\documentclass[aps,prd,nofootinbib,twocolumn,a4paper,superscriptaddress]{revtex4-1}
\usepackage{graphicx}
\usepackage{multirow}
\usepackage{latexsym}
\usepackage{hyperref}
\usepackage[british]{babel}
\usepackage{amsmath}
 \usepackage{color}
\begin{document}

\newcommand{\be}{\begin{equation}}
\newcommand{\ee}{\end{equation}}
\newcommand{\bq}{\begin{eqnarray}}
\newcommand{\eq}{\end{eqnarray}}
\newcommand{\bsq}{\begin{subequations}}
\newcommand{\esq}{\end{subequations}}
\newcommand{\bc}{\begin{center}}
\newcommand{\ec}{\end{center}}

\title{Semi-analytic calculation of cosmic microwave background anisotropies from wiggly and superconducting cosmic strings}

\author{I. Yu. Rybak}
\email[]{Ivan.Rybak@astro.up.pt}
\affiliation{Centro de Astrof\'{\i}sica da Universidade do Porto, Rua das Estrelas, 4150-762 Porto, Portugal}
\affiliation{Instituto de Astrof\'{\i}sica e Ci\^encias do Espa\c co, CAUP, Rua das Estrelas, 4150-762 Porto, Portugal}
\affiliation{Faculdade de Ci\^encias, Universidade do Porto, Rua do Campo Alegre 687, 4169-007 Porto, Portugal}
\author{A. Avgoustidis}
\email[]{Anastasios.Avgoustidis@nottingham.ac.uk}
\affiliation{School of Physics and Astronomy, University of Nottingham, University Park, Nottingham NG7 2RD, England}
\author{C. J. A. P. Martins}
\email{Carlos.Martins@astro.up.pt}
\affiliation{Centro de Astrof\'{\i}sica da Universidade do Porto, Rua das Estrelas, 4150-762 Porto, Portugal}
\affiliation{Instituto de Astrof\'{\i}sica e Ci\^encias do Espa\c co, CAUP, Rua das Estrelas, 4150-762 Porto, Portugal}

\date{13 August 2019}

\begin{abstract}
We study how the presence of world-sheet currents affects the evolution of cosmic string networks, and their impact on predictions for the cosmic microwave background (CMB) anisotropies generated by these networks. We provide a general description of string networks with currents and explicitly investigate in detail two physically motivated examples: wiggly and superconducting cosmic string networks. By using a modified version of the CMBact code, we show quantitatively how the relevant network parameters in both of these cases influence the predicted CMB signal. Our analysis suggests that previous studies have overestimated the amplitude of the anisotropies for wiggly strings. For superconducting strings the amplitude of the anisotropies depends on parameters which presently are not well known---but which can be measured in future high resolution numerical simulations.
\end{abstract}
\pacs{98.80.Cq, 11.27.+d, 98.80.Es}
\keywords{Cosmology, Topological defects, Domain walls, Numerical simulation, VOS model}
\maketitle

\section{Introduction}

As demonstrated in~\cite{Kibble}, symmetry breaking processes in early universe scenarios can lead to the formation of topologically stable line-like concentrations of energy, known as cosmic strings (for general reviews see~\cite{ShellardVilenkin,HindmarshKibble}). These one-dimensional objects evolve and interact with each other, forming a cosmic string network. Depending on their origin, strings can have significantly different properties and observational signatures. Examples of theoretically well-motivated scenarios where the presence of cosmic strings is expected include brane inflation~\cite{PolchinskiCopelandMyers,SarangiTye,FirouzjahiTye,JonesStoicaTye}, supersymmetric grand unified theories with hybrid inflation~\cite{JeannerotRocherSakellariadou,CuiMartinMorrisseyWells,JeannerotPostma,AchucarroCeli,DavisMajumdar,Allys} and many others~\cite{KoehnTrodden,SMASH,LazaridesPeddieVamvasakis,ChernoffTye}. In most cases, cosmic strings are stable and survive to the present era, acting as fossils for these models. Hence, quantitative bounds placed on string networks can lead to strong constraints on the underlying early universe model.

One difficulty is precisely that different models can produce strings with different properties, with varying observational predictions for the corresponding string networks. Hence, in order to achieve reliable observational constraints on the underlying early universe models from cosmic string network phenomenology, one needs to develop an accurate description of cosmic string network evolution, taking into account the distinctive features of different types of cosmic strings. One way to accomplish this task is through numerical simulations~\cite{Blanco-PilladoOlumShlaer,HindmarshLizarragaUrrestillaDaverioKunz}. This approach provides reliable results, but is currently limited by computer capabilities, especially when one tries to include non-trivial cosmic string features like world-sheet currents. At present, multi-tension cosmic string networks and strings with currents are very time-consuming to model, and cannot be simulated with both high-resolution and sufficiently large dynamic range (see~\cite{LizarragaUrrestilla} for a recent field theory simulation of $pq$-strings). Further, numerical simulations have to be repeated for different values of cosmological and string parameters and are thus not particularly flexible for parameter determination through direct confrontation with observational data. 

There is an alternative -- and largely complementary -- semi-analytic approach for the description of cosmic string network evolution based on the velocity-dependent one-scale (VOS) model~\cite{MartinsShellard,MartinsShellard2}. In this treatment it is much easier to add non-trivial features for cosmic strings~\cite{Martins,VieiraMartinsShellard,Vortons,OliveiraAvgoustidisMartins,AvgoustidisShellard,AvgoustidisShellard2,Copeland:2005cy,Tye:2005fn} allowing evolution over large dynamical ranges that cannot be achieved by numerical simulations. However, semi-analytic descriptions involve free parameters, which can only be reliably calibrated by comparison to simulations. As a result, a combination of such analytic descriptions and numerical simulations is at present the best approach for studying the evolution of cosmic string networks with non-trivial properties. 

Among different methods for detecting observational signals from cosmic string networks, cosmic microwave background (CMB) anisotropies offer one of the most sensitive and robust probes~\cite{Planck2013}. Current results obtained using cosmic string network simulations~\cite{Lazanau2,LizarragaUrrestillaDaverioHindmarshKunz} and calibrated semi-analytic descriptions~\cite{Tom} yield very similar constraints for simple global cosmic strings, the current limit on the string tension being at the level of $G \mu \lesssim 10^{-7}$. However, as discussed above, it is the latter approach that allows us to go beyond these vanilla strings and quantitatively study the observational effects of additional properties on cosmic strings.   

One such additional feature that we can anticipate in many cosmic string models is the presence of a worldsheet current. This can be caused by a coupling between the field forming the cosmic string and other fields, by trapped charged fermion modes along the string~\cite{Witten} (which is common in supersymmetric models~\cite{DavisDavisTrodden,DavisDavisTrodden2}), by trapped vector fluxes on non-Abelian strings~\cite{Everett}, and other specific mechanisms (for example symmetry breaking of an accidental symmetry in SU(2) stings~\cite{HindmarshRummukainenWeir}). From a phenomenological point of view, the presence of such currents gives rise to effective, macroscopic properties on the string. For example, small-scale structure (wiggles) on strings can be described by a specific type of current~\cite{Carter90,Carter95,Vilenkin}. 

In what follows we will show quantitatively how the presence of currents on the string worldsheet can affect observational predictions for the string CMB signal, paying particular attention to the special cases of wiggly strings and superconducting strings. In the case of wiggly strings this generalises and extends the work of~\cite{PogosianVachaspati}, where string wiggles were taken into account through a constant free parameter $\alpha$. In our approach we can construct the most general model for wiggly strings, leading to a full description of wiggles, including their evolution and their effect on the string equations of motion. For superconducting strings, some relevant model parameters are less well known (due to the lack of numerical simulations of these models) but we are also able to provide a full description. In both cases, our results will enable a more detailed and robust comparison to observations, which we leave for future work.

\section{String model with currents}

In order to obtain an effective two-dimensional Lagrangian of a string-like object from a four-dimensional field theory, one usually follows the procedure of~\cite{Witten}. This coarse-graining approach unavoidably involves the loss of some of the features of the original four-dimensional description; in particular, it cannot describe key properties of superconducting strings like current saturation and supersonic wiggle propagation. As a result, there is only a phenomenological approach to reproduce properties of the original four-dimensional model \cite{CarterPeter,CarterPeter2}. On the other hand, one is often interested in averaged equations of motion and these can be the same for different Lagrangians (for an explicit example see~\cite{Vortons}). Thus, focusing on deriving the exact form of the Lagrangian is not necessarily the most productive route to obtaining accurate string network evolution. 

Bearing in mind the subtleties described above, we consider the general form of a two-dimensional Lagrangian involving an arbitrary function of a string current. First, note that a current on a two-dimensional space can be represented as a derivative of a scalar field $\varphi$,
\begin{equation}
   \label{Current2d}
    J_a = \varphi_{,a},  
\end{equation}
where $_{,a}=\frac{\partial}{\partial \sigma^a}$, with $\sigma^a$ the coordinates on the string worldsheet (Latin indexes $a$, $b$ run over $0,1$).

We can thus build three possible terms ``living" on the worldsheet, out of which the Lagrangian will be constructed
\begin{equation}
\begin{gathered}
   \label{Terms}
   \text{[1]: } \; \varphi^{,a} \varphi^{,b} \gamma_{ab} = \kappa, \\ 
   \text{[2]: } \; \varepsilon^{ac} \varepsilon^{bd} \gamma_{ab} \gamma_{cd} = \gamma, \\ 
   \text{[3]: } \; \varepsilon^{ac} \varepsilon^{bd} \gamma_{ab} \varphi_{,c} \varphi_{,d} = \Delta,   
\end{gathered}
\end{equation}
where $\gamma_{ab}=g_{\mu \nu} x^{\mu}_{,a} x^{\nu}_{,b}$ is the induced metric on the string worldsheet ($g_{\mu \nu}$ being the background space-time metric with Greek indexes $\mu$, $\nu$ corresponding to 4-dimensional space-time coordinates), $\gamma$ is the determinant of the induced metric and $\varepsilon^{ab}$ is the Levi-Civita symbol in two dimensions.

The term $\Delta$ is motivated by the Dirac-Born-Infeld (DBI) action for cosmic strings (relevant studies can be found in \cite{Nielsen}, \cite{BabichevBraxCapriniMartinDanieleSteer} and \cite{OliveiraAvgoustidisMartins}). We would like to stress the fact that $\kappa$ and $\Delta$ are not independent variables. However, it is beneficial to introduce them as they help to organize the equations into useful form.

Taking into account the three possible terms in Eq.~(\ref{Terms}) we can write down the general form of the action generalising the Nambu-Goto action to the case of a string with current
\begin{equation}
   \label{ActionGeneral}
    S = - \mu_0 \int f(\kappa, \gamma, \Delta) \sqrt{-\gamma/2} d^2 \sigma,  
\end{equation}
where $\mu_0$ is a constant of dimensions $[Energy]^2$ determined by the symmetry breaking scale giving rise to string formation.

The arbitrary choice of the function $f(\kappa,\gamma,\Delta)$ can break reparametrisation invariance of the generalized action of Eq.~(\ref{ActionGeneral}). In order to preserve invariance of the action under reparametrizations, the last two terms of Eq.~\ref{Terms} should be connected in the following way $f(\kappa,\Delta/\gamma)$. Hereinafter  for the sake of simplicity the function $f(\kappa,\Delta/\gamma)$ in equations will be denoted just as $f$.

Assuming that cosmic strings are moving in a flat Friedmann-Lemaitre-Robertson-Walker  (FLRW) background with metric $ds^2 = a^2(\tau) \left( d\tau^2 - dl^2 \right)$, we can build the stress-energy tensor from the action~(\ref{ActionGeneral})
\begin{equation}
\begin{split}
   \label{StressEnergyGeneral}
    & T^{\mu \nu} (y) = \frac{\mu_0}{\sqrt{-g}} \int d^2 \sigma   \sqrt{-\gamma} \delta^{(4)}(y-x(\sigma))   \\
    & \left( \tilde{U} \tilde{u}^{\mu} \tilde{u}^{\nu} - \tilde{T} \tilde{v}^{\mu} \tilde{v}^{\nu}  - \Phi (\tilde{u}^{\mu} \tilde{v}^{\nu} + \tilde{v}^{\mu} \tilde{u}^{\nu}) \right),
\end{split}
\end{equation}
where $\tilde{u}^{\mu} = \frac{\sqrt{\epsilon} \dot{x}^{\mu}}{(-\gamma)^{1/4}}$ and $\tilde{v}^{\mu} = \frac{ x^{\prime \mu}}{\sqrt{\epsilon} (-\gamma)^{1/4}}$ are orthonormal timelike and spacelike  vectors respectively ($\tilde{u}^{\mu} \tilde{u}_{\mu}=1$, $\tilde{v}^{\mu} \tilde{v}_{\mu}=-1$), $\epsilon = \sqrt{\frac{\textbf{x}^{\prime \, 2}}{1-\dot{\textbf{x}}^2}}$ and
\begin{eqnarray}
   \label{U}
    & \tilde{U} =   f - 2\frac{\partial f}{\partial \gamma} \frac{\Delta}{\gamma} - 2  \gamma^{00} \frac{\partial f}{\partial \kappa} \dot{\varphi}^2 + 2  \gamma^{11} \frac{\partial f}{\partial \Delta} \varphi^{\prime \, 2}  , \\
    \label{T}
    & \tilde{T} =  f - 2\frac{\partial f}{\partial \gamma} \frac{\Delta}{\gamma} - 2   \gamma^{11} \frac{\partial f}{\partial \kappa} \varphi^{\prime \, 2} + 2  \gamma^{00} \frac{\partial f}{\partial \Delta} \dot{\varphi}^2 , \\
    \label{Phi}
    & \Phi =  \frac{2 }{\sqrt{-\gamma}} \left( - \frac{\partial f}{\partial \kappa} -\frac{\partial f}{\partial \Delta} \right) \varphi^{\prime} \dot{\varphi}\,;
\end{eqnarray}
here and henceforth dots and primes respectively denote time and space derivatives.

It is important to note that for this modification of the Lagrangian, the stress-energy tensor~(\ref{StressEnergyGeneral}) has non-diagonal terms induced by the presence of the current. Let us obtain the equations of motion for the action~(\ref{ActionGeneral}) using the definitions of $\tilde{U}$ in~(\ref{U}), $\tilde{T}$ in~(\ref{T}) and $\Phi$ in~(\ref{Phi}). A variation of the action~(\ref{ActionGeneral}) with respect to $x^{\mu}$ and $\varphi$ gives
\begin{eqnarray}
   \label{EquationMotionGeneral1}
    & \partial_{\tau} ( \epsilon \tilde{U} ) + \frac{\dot{a}}{a} \epsilon \left( \dot{\textbf{x}}^2 (\tilde{U} + \tilde{T}) + \tilde{U} - \tilde{T} \right) =  \partial_{\sigma} \Phi,  \\
    & \ddot{\textbf{x}} \epsilon \tilde{U} + \dot{\textbf{x}} \epsilon \frac{\dot{a}}{a} \left( 1-\dot{\textbf{x}}^2 \right) \left( \tilde{U} + \tilde{T} \right) = \nonumber \\
    \label{EquationMotionGeneral2}
    & =\partial_{\sigma} \left( \frac{ \tilde{T} }{\epsilon} \textbf{x}^{\prime} \right) +  \textbf{x}^{\prime} \left( 2 \frac{\dot{a}}{a} \Phi + \dot{\Phi} \right) + 2 \Phi \dot{\textbf{x}}^{\prime},\\
    \label{EquationMotionGeneral3}
    & \partial_{\tau} \left(   \left( \frac{\partial f}{\partial \kappa} + \frac{\partial f}{\partial \Delta} \right)  \epsilon \dot{\varphi} \right) = \partial_{\sigma} \left(  \left( \frac{\partial f}{\partial \kappa} + \frac{\partial f}{\partial \Delta} \right)  \frac{\varphi^{\prime}}{\epsilon} \right).
\end{eqnarray}
where we have chosen a parametrisation satisfying the transverse temporal conditions $\dot{\textbf{x}} \cdot \textbf{x}^{\prime}=0$ and $x^0=\tau$. 

As can be seen from the equations of motion~(\ref{EquationMotionGeneral1}) and~(\ref{EquationMotionGeneral2}), string dynamics does not depend explicitly on the form of the current contribution $f(\kappa, \Delta/\gamma)$. The dynamics of the string is defined completely by $\tilde{U}$, $\tilde{T}$ and $\Phi$, which can be associated to mass per unit length and string tension. Indeed, it is only the dynamics of $\varphi$ itself -- Eq.~(\ref{EquationMotionGeneral3}) -- that explicitly depends on $\partial f/\partial\kappa$ and $\partial f/\partial\Delta$. This provides us an alternative approach to studying string dynamics effectively, without an explicit connection between an effective Nambu-Goto-like action and the original field theory model. One can instead study the behaviour of $\tilde{U}$, $\tilde{T}$ and $\Phi$ in the original four-dimensional model in the framework of field theory (as it was done for example in~\cite{Ringeval1,LilleyPeteMartin,Ringeval2,LilleyMarcoMartinPeter,Allys2}) and then insert the dynamics of $\tilde{U}$, $\tilde{T}$ and $\Phi$ in the equations of motion (\ref{EquationMotionGeneral1}) and~(\ref{EquationMotionGeneral2}).

Additionally, we also note that one can easily generalize the equations of motion~(\ref{EquationMotionGeneral1})-(\ref{EquationMotionGeneral3}) to include any number of uncoupled scalar fields, associated to corresponding currents. In this case, we can simply rewrite the variables $\kappa$ and $\Delta$ as
\begin{equation}
   \label{kappa}
    \kappa_i =  \gamma_{ab} \varphi_i^{,a} \varphi_i^{,b},  \quad \Delta_i = \varepsilon^{ac} \varepsilon^{bd} \gamma_{ab} \varphi_{i,c} \varphi_{i,d}\,,
\end{equation}
where the index $i$ runs over the number of fields. There is no summation over $i$; if a sum over this index is to be taken it will be written explicitly.

Definitions~(\ref{U}),~(\ref{T}) and~(\ref{Phi}) in the case of multiple currents generalise to
\begin{eqnarray}
   \label{U2}
      & \hspace{-0.1in} \tilde{U}= f +   2 \sum_i  \left( -\gamma^{00} \frac{\partial f}{\partial \kappa_i} \dot{\varphi_i}^2 + \gamma^{11} \frac{\partial f}{\partial \Delta_i} \varphi_i^{\prime \, 2} - \frac{\partial f}{\partial \gamma} \frac{\Delta_i}{\gamma} \right), \\
    \label{T2}
   &  \hspace{-0.1in} \tilde{T} =  f + 2 \sum_i \left( -\gamma^{11} \frac{\partial f}{\partial \kappa_i} \varphi_i^{\prime \, 2} +  \gamma^{00} \frac{\partial f}{\partial \Delta_i} \dot{\varphi_i}^2 - \frac{\partial f}{\partial \gamma} \frac{\Delta_i}{\gamma} \right) , \\
    \label{Phi2}
   & \Phi = \frac{2}{\sqrt{-\gamma}} \sum_i \left(- \frac{\partial f}{\partial \kappa_i} -\frac{\partial f}{\partial \Delta_i} \right)  \varphi_i^{\prime} \dot{\varphi}_i .
\end{eqnarray}

With definitions~(\ref{U2})-(\ref{Phi2}) the form of the stress-energy tensor~(\ref{StressEnergyGeneral}) and the equations of motion~(\ref{EquationMotionGeneral1})-(\ref{EquationMotionGeneral2}) stay unchanged. On the other hand, the equation of motion for the scalar field~(\ref{EquationMotionGeneral3}) is substituted by the set of equations
\begin{eqnarray}
    \label{EquationMotionGeneral32}
     & \partial_{\tau} \left(  \left( \frac{\partial f}{\partial \kappa_i} + \frac{\partial f}{\partial \Delta_i} \right) \epsilon \dot{\varphi}_i \right) = \partial_{\sigma} \left( \left( \frac{\partial f}{\partial \kappa_i} + \frac{\partial f}{\partial \Delta_i} \right)  \frac{\varphi_i^{\prime}}{\epsilon} \right).
\end{eqnarray}

We see, therefore, that if we extend the action~(\ref{ActionGeneral}) to include additional scalar fields $\varphi_i$, the structure of the equations of motion together with the form of the general stress-energy tensor remains unchanged; we only need to add a new index $i$ to $\kappa$ and $\Delta$. This fact will be useful in our considerations below. For now, let us diagonalise the stress-energy tensor~(\ref{StressEnergyGeneral}) and define the \emph{mass per unit length} and \emph{tension} for these strings with currents, following Refs.~\cite{Carter90,LilleyPeteMartin}
\begin{eqnarray}
\label{Ueq}
    & T^{\mu}_{\nu} u^{\nu} = U \delta^{\mu}_{\nu} u^{\nu}, \\
\label{Teq}
    & T^{\mu}_{\nu} v^{\nu} = T \delta^{\mu}_{\nu} v^{\nu}\,.
\end{eqnarray}
The new orthonormal timelike $u^\mu$ and spacelike $v^\mu$ vectors are eigenvectors of the stress-energy tensor~(\ref{StressEnergyGeneral}) with corresponding eigenvalues $U$ (mass per unit length) and $T$ (tension). These eigenvalues are related to the original $\tilde{U}$, $\tilde{T}$ and $\Phi$ in~(\ref{U})-(\ref{Phi}) by
\begin{eqnarray}
\label{MassPerUnitLength}
   & U = \mu_0/2 \left( \tilde{U} + \tilde{T} + \Delta \right), \\
\label{Tension}
   & T = \mu_0/2 \left(\tilde{U} + \tilde{T} - \Delta \right)\,,
\end{eqnarray}
while the eigenvectors can be expressed in terms of the original $\tilde{u}^{\mu}$ and $\tilde{v}^{\mu}$ as 
\begin{eqnarray}
\label{TimeLike}
   & u^{\mu} = a \tilde{u}^{\mu} + \sqrt{a^2 - 1} \tilde{v}^{\mu}, \\
\label{SpaceLike}
   & v^{\mu} = \sqrt{a^2 - 1} \tilde{u}^{\mu} + a \tilde{v}^{\mu},
\end{eqnarray}
with 
$$a=\frac{1}{2} \left[ 1+ \frac{\tilde{U}-\tilde{T}}{\Delta} \right]$$ and $$\Delta=\sqrt{(\tilde{U}-\tilde{T})^2 - 4 \Phi^2}\,.$$

The passage from the equations of motion of a single string segment to an effective description of a whole network of strings is done through an averaging procedure~\cite{MartinsShellard} leading to the VOS model for cosmic strings. Following this approach, we begin by dotting equation~(\ref{EquationMotionGeneral2}) with vectors $\dot{\textbf{x}}$ and $\textbf{x}^{\prime}$ and using the property of our parametrization 
$\dot{\textbf{x}} \cdot \textbf{x}^{\prime}=0$ to obtain 
\begin{eqnarray}
    &  \dot{\textbf{x}} \cdot \ddot{\textbf{x}} \epsilon \tilde{U} + \dot{\textbf{x}}^2 \epsilon \frac{\dot{a}}{a} \left( 1-\dot{\textbf{x}}^2 \right) \left( \tilde{U} + \tilde{T} \right) = \nonumber \\
    \label{EquationMotionGeneral21}
    & =\frac{ \tilde{T} }{\epsilon} \dot{\textbf{x}} \cdot \textbf{x}^{\prime \prime} - 2 \Phi \ddot{\textbf{x}} \cdot \textbf{x}^{\prime}, \\
    &  \textbf{x}^{\prime} \cdot \ddot{\textbf{x}} \epsilon \tilde{U} - \textbf{x}^{\prime} \cdot \textbf{x}^{\prime \prime} \frac{ \tilde{T} }{\epsilon} +  2 \Phi \textbf{x}^{\prime \prime} \cdot \dot{\textbf{x}}   = \nonumber \\
    \label{EquationMotionGeneral22}
    & =  \textbf{x}^{\prime 2} \left( 2 \frac{\dot{a}}{a} \Phi + \dot{\Phi} +  \frac{ \tilde{T^{\prime}} }{\epsilon} - \frac{ \tilde{T} }{\epsilon^{2}} \epsilon^{\prime} \right).
\end{eqnarray}

Using the expression $\frac{\epsilon^{\prime}}{\epsilon} = \frac{ \textbf{x}^{\prime} \cdot \textbf{x}^{\prime \prime}}{ \textbf{x}^{\prime 2} } - \frac{ \textbf{x}^{\prime} \cdot \ddot{\textbf{x}} }{1-\dot{\textbf{x}}^{2}}$ we can eliminate the terms proportional to $\epsilon^{\prime}$ and 
$\textbf{x}^{\prime} \cdot \ddot{\textbf{x}}$ obtaining the equation
\begin{equation}
\begin{gathered}
      \dot{\textbf{x}} \cdot \ddot{\textbf{x}} \epsilon \tilde{U} + \dot{\textbf{x}}^2 \epsilon \frac{\dot{a}}{a} \left( 1-\dot{\textbf{x}}^2 \right) \left( \tilde{U} + \tilde{T} \right) - \frac{ \tilde{T} }{\epsilon} \dot{\textbf{x}} \cdot \textbf{x}^{\prime \prime} =  \\
    \label{EquationMotionGeneral222}
     = 2 \Phi \frac{1-\dot{\textbf{x}}^2}{\tilde{U}-\tilde{T}} \left( \tilde{T}^{\prime} + \epsilon \left( 2 \frac{\dot{a}}{a} \Phi + \dot{\Phi} - 2 \Phi \frac{\textbf{x}^{\prime \prime} \cdot \dot{\textbf{x}}}{ \textbf{x}^{\prime 2} } \right) \right).
\end{gathered}
\end{equation}

We now introduce the macroscopic variables 
\begin{eqnarray}
   \label{MacroVarE}
   & E = \mu_0 a \int \tilde{U} \epsilon d \sigma\,,\\  
      \label{MacroVarEo}
   & E_0 = \mu_0 a \int \epsilon d \sigma\,,\\
   \label{MacroVarV}
 & v^2 = \left\langle \dot{x}^2 \right\rangle\,,
\end{eqnarray}
where $\left\langle ... \right\rangle = \frac{\int ...\,  \epsilon d \sigma}{\int  \epsilon d \sigma}$ denotes the (energy-weighted) 
averaging operation. These macroscopic quantities are, respectively, the total energy, the 'bare' energy (without the contribution form the current) and the Root-Mean Squared (RMS) velocity. Using these definitions we proceed to average equations~(\ref{EquationMotionGeneral222}) and~(\ref{EquationMotionGeneral1}) finding
\begin{eqnarray}
   \label{AverEquationMotionGeneral1}
    & \dot{E} + \frac{\dot{a}}{a} E \left( \upsilon^2 \left( 1 +  W \right)  -  W \right) = \left\langle \Phi^{\prime}/\epsilon \right\rangle E_0 ,  \\
     & \dot{\upsilon}  + \upsilon \frac{\dot{a}}{a} \left( 1-\upsilon^2 \right) \left( 1 +  W  \right) - (1-\upsilon^2) W \frac{k(\upsilon)}{R_c}  \nonumber =  \\
    \label{AverEquationMotionGeneral2}
    & = \left< 2 \frac{\Phi}{\tilde{U}} \frac{1-\dot{\textbf{x}}^2}{1-\tilde{T}/\tilde{U}} \left(\frac{ \tilde{T}^{\prime}}{\epsilon} + 2 \frac{\dot{a}}{a} \Phi + \dot{\Phi} - 2 \Phi \frac{\textbf{x}^{\prime \prime} \cdot \dot{\textbf{x}}}{ \textbf{x}^{\prime 2} } \right)\right>.
\end{eqnarray}
Here, we have defined $W=\left\langle \tilde{T}/\tilde{U} \right\rangle$ and introduced the average comoving radius of curvature of strings in the network, $R_c$, and the curvature parameter, $k(\upsilon)$, satisfying $\left< \frac{\dot{\textbf{x}}}{\epsilon} \cdot \left( \frac{\textbf{x}^{\prime}}{\epsilon} \right)^{\prime} \right>= \frac{k(\upsilon)}{R_c} \upsilon (1-\upsilon^2)$.  For ordinary cosmic strings, an accurate ansatz for the curvature parameter as a function of velocity 
\begin{equation}
   \label{MomentumParameter}
  k(\upsilon) = \frac{2 \sqrt{2}}{\pi} (1-v^2) (1+2 \sqrt{2}v^3) \frac{1-8v^6}{1+8v^6} \,,
\end{equation}
has been derived in~\cite{MartinsShellard2}. We assume that this function stays valid for strings with currents as well.

Following the procedure of~\cite{MartinsShellard,MartinsShellard2}, we rewrite the averaged equations of motion~(\ref{AverEquationMotionGeneral1}-\ref{AverEquationMotionGeneral2}) in terms of more convenient macroscopic variables: the comoving characteristic length $L_c$ and the comoving correlation length $\xi_c$, which are related to the energies in (\ref{MacroVarE}-\ref{MacroVarEo}) by the following expressions 
$$E=\frac{\mu_0 V}{a^2 L_c^2}$$ 
and 
$$E_0=\frac{\mu_0 V}{a^2 \xi_c^2}\,,$$
where $V$ is the volume over which the averaging has been performed. In addition, we employ the VOS model approximation that the average radius of curvature of cosmic strings in the network is equal to the correlation length, i.e. $R_c \approx \xi_c$.  Assuming further that the averaged macroscopic quantities can be split as $\left< \Phi \tilde{U} \tilde{T} \right> = \left< \Phi \right> \left< \tilde{U} \right> \left< \tilde{T} \right> $ we obtain the following system of equations
\begin{eqnarray}
   \label{AvEquationMotionGeneral1}
    & 2 \dot{L_c} = \frac{\dot{a}}{a} L_c \left( \upsilon^2 \left( 1 +  W \right)  -  W +1 \right) -  \frac{\sqrt{1-\upsilon^2} Q_{,s}}{\hat{U}} ,  \\
     & \dot{\upsilon}  + \upsilon \frac{\dot{a}}{a} \left( 1-\upsilon^2 \right) \left( 1 +  W  \right) - (1-\upsilon^2) W \frac{k(\upsilon)}{\xi_c}   \nonumber =  \\
    \label{AvEquationMotionGeneral2}
    & \hspace{-0.1in} = 2 \frac{ Q }{\hat{U} } \frac{1-\upsilon^2}{1-W} \left(\sqrt{1-\upsilon^2}\, \hat{T}_{,s} + \dot{Q} + 2 Q \left( \frac{\dot{a}}{a} - \frac{k(\upsilon) \upsilon}{\xi_c}  \right) \right),
\end{eqnarray}
where $\left< \Phi \right> = Q$, $\left< \dot{\Phi} \right> = \dot{Q}$, $\hat{U}= \left< \tilde{U} \right>$, $\hat{T}= \left< \tilde{T} \right>$, the correlation and characteristic lengths are related by $\xi_c = L_c \sqrt{\hat{U}}$ and a new derivative variable $_{,s} = \frac{\partial}{\partial s}$ has been introduced, corresponding to the parametrization $ds=\sqrt{\textbf{x}^{\prime 2}}d\sigma$. 

Equations~(\ref{AvEquationMotionGeneral1}-\ref{AvEquationMotionGeneral2}) are the averaged macroscopic equations describing a network of cosmic strings with a current. It is apparent that scaling solutions ($L_c = \varepsilon_c \tau, \; \upsilon = \text{const}$ when $a \propto \tau^n$ with $\varepsilon_c$ and $n$ constants~\cite{MartinsShellard,MartinsShellard2}) exist if the averaged quantities $\hat{U}$, $\hat{T}$ and $Q$ are appropriately restricted. In particular we see from~(\ref{AvEquationMotionGeneral1})-(\ref{AvEquationMotionGeneral2}) that scaling behaviour -- typical for ordinary string networks -- can arise when $\hat{U}, \hat{T}, Q = \text{const}$, while $T_{,s} \; \text{and} \; Q_{,s} \sim 1/\tau$. Additionally, the requirement of a well-defined $\varepsilon_c$ implies the condition
\begin{equation}
   \label{W-restriction}
     \upsilon^2 < \frac{2+n(W-1)}{n(W+1)}.
\end{equation}

Equation~(\ref{W-restriction}) relates the rms string velocity $\upsilon$ to the ratio $W=\left\langle \tilde{T}/\tilde{U} \right\rangle$ for a given expansion rate (characterisd by $n$) for a cosmic string network with currents. These general relations will be useful when we consider the special case of a wiggly string network. 

We now concentrate on how these modifications can influence predictions for the CMB anisotropy from cosmic strings. We follow the approach of~\cite{PogosianVachaspati,TurokPenSeljak,AlbrechtBattyeRobinson}. Rather than working with the full network of cosmic strings, we consider a number of straight string segments in Minkowski space that decay according to the evolution of strings in an expanding FLRW metric, and have velocities and lengths determined by the VOS model.

We start from the Fourier transform of the stress-energy tensor~(\ref{StressEnergyGeneral}) of a single straight string segment on which the  contribution from string currents has been averaged as above 
\begin{equation}
   \label{Stress-EnergyGeneralFour}
   \begin{split}
    & \Theta^{\mu \nu} = \mu_0 \int_{- \xi_0 \tau /2}^{ \xi_0 \tau /2}  \biggl[ \hat{U} \epsilon   \dot X^{\mu} \dot X^{\nu}- \hat{T} \frac{X^{\prime \mu} X^{\prime \nu}}{\epsilon} - \\
    & \quad - Q \left(\dot X^{\mu} X^{\prime \nu}+ \dot X^{\nu} X^{\prime \mu} \right)  \biggl] \text{e}^{i {\bf k}\cdot {\bf X}} d \sigma,
\end{split}
\end{equation}
where the vector $X^{\mu} = x_0^{\mu}+\sigma X^{\prime \, \mu} + \tau \dot X^{\mu}$ represents the straight, stick-like solution for a string moving with velocity $\upsilon$ (so that $\dot X^{\mu} \dot X_{\mu}=1-\upsilon^2$) and with worldsheet coordinates $\sigma$ and $\tau$ in the transverse temporal gauge. The  comoving length of a string segment at conformal time $\tau$ is $\xi_0 \tau$, where $\xi_0$ will be determined from the macroscopic evolution equations (\ref{AvEquationMotionGeneral1}-\ref{AvEquationMotionGeneral2}). Variables $\hat{U}$, $\hat{T}$ and $Q$ are constants for the straight string, as follows from the equations of motion~(\ref{EquationMotionGeneral1}-\ref{EquationMotionGeneral3}). The four-vector $x_0^{\mu}=(1,{\bf x_0})$ is a random location for a single string segment, while $X^{\prime\mu}$ and $\dot X^{\mu}$ are randomly oriented and satisfy the transverse condition $X^{\prime}_{\mu} \dot X^{\mu}=0$. We can choose these vectors\footnote{Note that although we work in the transverse temporal gauge we have chosen the normalization ${\bf X}^{\prime 2}=1$. This may seem to be inconsistent as ${\bf X}^{\prime 2}=\epsilon^2 (1-\dot {\bf X}^2)$ and $\epsilon$ is evolving according to equation (\ref{EquationMotionGeneral1}). However, we are implicitly taking this effect into account by having the limits of the integral (\ref{Stress-EnergyGeneralFour}) be time-dependent through the time evolution of $\xi_0$. This evolves according to the macroscopic equation (\ref{AvEquationMotionGeneral1}), which has been derived by averaging equations~(\ref{EquationMotionGeneral222}) and~(\ref{EquationMotionGeneral1}). } as 
\begin{eqnarray}
\label{UnitVectors}
     & \dot X^{\mu} = \begin{pmatrix} 1 \\ \upsilon ( \cos \theta \cos \phi \cos \psi - \sin \phi \sin \psi) \\  \upsilon ( \cos \theta \sin \phi \cos \psi + \cos \phi \sin \psi ) \\ -\upsilon \sin \theta \cos \psi \end{pmatrix}, \\
     & X^{\prime \; \mu} = \begin{pmatrix} 0 \\ \sin \theta \cos \phi \\  \sin \theta \sin \phi \\ \cos \theta \end{pmatrix}\,. \\ \nonumber
\end{eqnarray}

Without loss of generality we can choose the wave vector along the third axis  $\textbf{k}=k \hat{k}_3$ and integrating over $\sigma$ we obtain the following expressions
\begin{eqnarray}
   \label{Stress-Energy00}
    & \Theta_{00} = \frac{\mu_0 \hat{U}}{\sqrt{1-v^2}} \frac{\sin(k X_3 \xi_0 \tau /2)}{k X_3/2} \cos(\textbf{k} \cdot \textbf{x}_0+k X_3 v \tau) , \\
   \label{Stress-EnergyIJ}    
    & \Theta_{ij} = \Theta_{00} \biggl[ v^2 \dot X_i \dot X_j - \hat{T} / \hat{U}(1-v^2) X_i^{\prime} X_j^{\prime} -   \nonumber \\ 
    &  - v Q/\hat{U} \left(\dot X_i X^{\prime}_j+ \dot X_j X^{\prime}_i \right) \biggl],
\end{eqnarray}
where the indices $i$, $j$ run over the 3-dimensional spatial coordinates.

The scalar, vector and tensor components can be defined as
\begin{equation}
\begin{gathered}
   \label{Scalar}
     \Theta^S = \left( 2 \Theta_{33} - \Theta_{11} - \Theta_{22} \right)/2 ,
 \end{gathered}
\end{equation}
\begin{equation}
\begin{gathered}
   \label{Vector}    
      \Theta^V = \Theta_{13},
 \end{gathered}
\end{equation}
\begin{equation}
\begin{gathered}
       \label{Tensor}    
      \Theta^T=\Theta_{12}.
 \end{gathered}
\end{equation}
Substituting~(\ref{Stress-Energy00}) and~(\ref{Stress-EnergyIJ}) in~(\ref{Scalar})-(\ref{Tensor}), we obtain the scalar, vector and tensor contributions for a straight string segment with stress-energy tensor~(\ref{StressEnergyGeneral}), (\ref{Stress-EnergyGeneralFour})
\begin{equation}
\begin{gathered}
   \label{Stress-EnergyChSc}
     \frac{2 \Theta^S}{\Theta_{00}} =\biggl[ v^2  (3 \dot{X}_3 \dot{X}_3 - 1) - 6 v Q / \hat{U} X_3^{\prime} \dot{X}_3 -   \\ 
       -(1-v^2) \hat{T} / \hat{U} (3 X_3^{\prime}  X_3^{\prime} -1) \biggl] ,
   \end{gathered}
\end{equation}
\begin{equation}
\begin{gathered}
   \label{Stress-EnergyChVect}    
      \frac{\Theta^V}{\Theta_{00}} =\biggl[ v^2  \dot{X}_1 \dot{X}_3 - \hat{T}/\hat{U} (1-v^2) X_1^{\prime}  X_3^{\prime} -  \\ 
     - v Q / \hat{U}  \left( X_1^{\prime} \dot{X}_3 + \dot{X}_1 X_3^{\prime}  \right) \biggl],
    \end{gathered}
\end{equation}
  \begin{equation}
\begin{gathered}
       \label{Stress-EnergyChTens}    
      \frac{\Theta^T}{\Theta_{00}} =\biggl[ v^2  \dot{X}_1 \dot{X}_2 - \hat{T}/\hat{U} (1-v^2) X_1^{\prime}  X_2^{\prime} - \\ 
     - v Q / \hat{U}  \left( X_1^{\prime} \dot{X}_2 + \dot{X}_1 X_2^{\prime}  \right) \biggl]  .
   \end{gathered}
\end{equation}

Following the prescription of reference~\cite{ACMS}, we can then calculate the unequal time two-point correlators by averaging over locations, string orientations and velocity orientations of the string segment

\begin{widetext}
\begin{equation}
   \label{Correlator}
    \left\langle \Theta^{I}(k, \tau_1) \Theta^{J}(k, \tau_2) \right\rangle = \frac{2 \mu_0^2 \mathcal{F}(\tau_1, \tau_2, \xi_0 )}{16 \pi^3} \int_0^{2 \pi} d \phi \int_0^{\pi} \sin \theta d \theta \int_0^{2 \pi} d \psi \int_0^{2 \pi} d \chi\, \Theta^{I}(k,\tau_1) \Theta^{J}(k, \tau_2)\,.
\end{equation}
\end{widetext}
Here, the indices $I$ and $J$ correspond to the scalar, vector, tensor and ``00" components. The function $\mathcal{F}(\tau_1, \tau_2, \xi_0)$ describes the string decay rate. It is chosen to have the same form as for ordinary (without currents) cosmic strings \cite{PogosianVachaspati} 
\begin{equation}
   \label{StringDecay}
     \mathcal{F} (\tau_1, \tau_2, \xi_0) = \frac{1}{\left( \xi_0 Max(\tau_1, \tau_2) \right)^3},
\end{equation}
but here $\xi_0$ is determined by the modified VOS equations~(\ref{AverEquationMotionGeneral1}-\ref{AverEquationMotionGeneral2}).
The phase $\chi = \textbf{k} \cdot \textbf{x}_0 $ arises from varying over string locations ${\bf x}_0$ (refer to equation (\ref{Stress-Energy00})), which we integrate over.

We can write the general form of the correlators as
\begin{eqnarray}
   \label{CorrelatorsForm}
    & \hspace{-0.1in} \left\langle \Theta^{I}(k, \tau_1) \Theta^{J}(k, \tau_2) \right\rangle = \frac{\mu_0^2 \mathcal{F}(\tau_1, \tau_2, \xi_0) }{k^2 (1-\upsilon^2)} B^{I-J}(\tau_1,\tau_2).
\end{eqnarray}
If we are only interested in the approximation $k \tau<1$ (superhorizon scales), we can expand $B^{I-J}$ keeping only terms that are up to $k^2$. In this case the non-zero correlators are the following 
\begin{eqnarray}
   \label{B001}
    & B^{00-00}(\tau_1,\tau_2) \approx \hat{U}^2 \xi_0^2 k^2 \tau_1 \tau_2, \\
    \label{BS1}
    & B^{S-S} \approx \frac{1}{5} \frac{B^{00-00}}{\hat{U}^2}(\tau_1,\tau_2) \times  \\
    &  \biggl( \hat{U}^2 \upsilon^4 + \hat{T} \hat{U} \upsilon^2 (1-\upsilon^2)+  \hat{T}^2 (1- \upsilon^2)^2 + 3 \upsilon^2 Q^2 \biggl), \nonumber \\
    \label{BV1}
    & B^{V-V} \approx \frac{1}{3} B^{S-S},\\
    \label{BT1}    
    &  B^{T-T} \approx \frac{1}{3} B^{S-S}.
\end{eqnarray}
In the Appendix we give exact expressions for the equal time two-point correlators $B^{I-J}(\tau)$ and provide semi-analytic expressions for the unequal time two-point correlators valid for all (i.e. from subhorizon through to superhorizon) modes $k$.

Having computed the correlators~(\ref{CorrelatorsForm}), let us now assume that the cosmic string network under consideration has reached a scaling regime. We can then assume that $\xi_0$, $\upsilon$ together with $\hat{U}$, $\hat{T}$ and $Q$ do not depend on $\tau$ and $\sigma$. To obtain an analytic estimate of the string-induced CMB anisotropy, let us consider the string network evolving in the matter domination epoch ($n=2$). For this case we can use the following solution of the linearised Einstein-Boltzmann equations~\cite{TurokPenSeljak,PenSpergelTurok}
\begin{equation}
 \begin{gathered} 
 \label{Pert1}
  \frac{\delta T}{T} = -\frac{1}{2} \int_{\tau_i}^{\tau_f} d\tau \dot{h}_{ij} n^i n^j ,\\
 \dot{h}_{ij} = \dot{h}_{ij}^{S} + \dot{h}_{ij}^{V} + \dot{h}_{ij}^{T} , 
\end{gathered}
\end{equation} 
\begin{equation}
 \begin{gathered}
 \label{PertS}
 \dot{h}_{ij}^{S} = -\rho \sum_k \text{e}^{i \textbf{k}\cdot \textbf{x}} \int_0^{\tau} d \tau^{\prime} \\
    \left(\frac{1}{3} \delta_{ij} \left( \frac{\tau^{\prime}}{\tau} \right)^6 (\Theta^{Tr} + 2 \Theta^S) - k_i k_j \left( \frac{\tau'}{\tau} \right)^4 \Theta^S \right),  \\
\end{gathered}
\end{equation}     
    \begin{equation}
 \begin{gathered}
 \label{PertV}
  \dot{h}_{ij}^{V} = \sum_k \text{e}^{i \textbf{k} \cdot \textbf{x}} \left( \dot{V}_{i} k_j + \dot{V}_{j} k_i \right),\\
  \dot{V}_{i} = \rho \int_0^{\tau} d \tau^{\prime} \left( \frac{\tau^{\prime}}{\tau} \right) \Theta^V_i, 
\end{gathered}
\end{equation} 
\begin{equation}
 \begin{gathered}
 \label{PertT}
 \dot{h}_{ij}^{T}=\rho \int_0^{\tau} d \tau^{\prime} k^3 \tau^{\prime 4} F(k \tau^{\prime},k \tau) \Theta_{ij}^T, \\
 F(k \tau^{\prime},k \tau) = G_1(k \tau^{\prime}) \dot{G}_2(k \tau) - G_2(k \tau^{\prime}) \dot{G}_1(k \tau),
\end{gathered}
\end{equation} 
where $\rho=16 \pi G$, $G_1(k \tau) = \frac{\cos(k \tau)}{(k \tau)^2}+\frac{\cos(k \tau)}{(k \tau)^3}$, $G_2(k \tau) = \frac{\cos(k \tau)}{(k \tau)^3}+\frac{\sin(k \tau)}{(k \tau)^2}$,  $\frac{\delta T}{T}$ are the CMB temperature fluctuations, $n^i$ is a unit vector defining the direction of CMB photons, and $\Theta^{Tr}$ is the trace of the Fourier transformed stress-energy tensor.

We can now compute the angular power spectrum $C_l$ of the CMB anisotropy using the expressions~\cite{TurokPenSeljak}:
\begin{equation}
 \begin{gathered} 
 \label{Cls}
  C_l^{S} = \frac{1}{2 \pi} \int_0^{\infty} k^2 dk  \\
 \left< \int_0^{\tau_0} d \tau  \left(\frac{1}{3} \dot{h}_1 + \dot{h}_2 \frac{d^2}{d(k \Delta \tau)^2}  \right) j_l(k \Delta \tau) \right>^2, 
\end{gathered}
\end{equation} 
 \begin{equation}
 \begin{gathered}  
 \label{Clv}
   C_l^{V}=\frac{2}{\pi} \int_0^{\infty} k^2 dk l(l+1)  \\
 \left< \int_0^{\tau0} d \tau \dot{h}^V \frac{d}{d(k \Delta \tau)} \left( j_l(k \Delta \tau) / (k \Delta \tau) \right) \right>^2,
\end{gathered}
\end{equation}  
 \begin{equation}
 \begin{gathered} 
 \label{Clt}
   C_l^{T} = \frac{1}{2 \pi} \int_0^{\infty} k^2 dk \frac{(l+2)!}{(l-2)!}  \\
  \left< \int_0^{\tau0} \frac{d \tau}{(k \Delta \tau)^2} \dot{h}^T  j_l(k \Delta \tau) \right>^2,
\end{gathered}
\end{equation} 
where $\Delta \tau = \tau_0 - \tau$ (with $\tau_0$ the value of conformal time today), $j_l(k \Delta \tau)$ are spherical Bessel functions, and $\dot{h}_1$, $\dot{h}_2$ are defined as
\begin{eqnarray}
 \label{h1}
&  \dot{h}_1(\tau) = - \rho \int d \tau^{\prime} \left(\frac{ \tau^{\prime} }{\tau} \right)^6 (\Theta^{Tr}(\tau^{\prime}) + 2 \Theta^S(\tau^{\prime})),  \\
 \label{h2}
& \dot{h}_2(\tau) = - \rho \int d \tau^{\prime}\left(\frac{ \tau^{\prime} }{\tau} \right)^4 \Theta^S(\tau^{\prime}).
\end{eqnarray}

We proceed by making a further approximation on the correlators~(\ref{CorrelatorsForm}). The dominant contribution to the two-point correlator is when $\tau_1 \rightarrow \tau_2$ (see for example~\cite{ACMS}), which allows us to approximate~(\ref{CorrelatorsForm}) as
\begin{equation}
\begin{gathered}
   \label{CorrelatorsForm2}
    \left\langle \Theta^{I}(k, \tau_1) \Theta^{J}(k, \tau_2) \right\rangle = \frac{\mu_0^2 \mathcal{F}(\tau_1, \tau_2, \xi_0) }{k^2 (1-\upsilon^2)} \times  \\
    B^{I-J}(\tau_1) \delta(\tau_1 - \tau_2),
\end{gathered}
\end{equation}
where $\delta(\tau_1-\tau_2)$ is Dirac delta function and $B^{I-J}(\tau_1) = B^{I-J}(\tau_1, \tau_1)$.

By using this form of the correlators~(\ref{CorrelatorsForm2}) one can rewrite equations~(\ref{Cls}),~(\ref{Clv}) and~(\ref{Clt}) as

\begin{widetext}
\begin{eqnarray}
 \label{Cls2}
&  C_l^{S} = \frac{\kappa^2}{2 \pi} \int_0^{ \infty } k^2 dk \int_0^{\tau0} d \tau_1 \int_0^{\tau0} d \tau_2  \int_0^{\tau_1} d \tau_1^{\prime}  \frac{f(\tau_1^{\prime},\varepsilon)}{k^2 (1-v^2)} \frac{\tau_1^{\prime 8}}{\tau_1^4 \tau_2^4} F_{sc} (\tau_1^{\prime})  , \\
 \label{Clv2}
&   C_l^{V}=\frac{2 \kappa^2}{\pi} \int_0^{\infty} k^2 dk l(l+1) \int_0^{\tau0} d \tau_1 \int_0^{\tau0} d \tau_2  \frac{j_l^{\prime}(k \tau_1)}{k \tau_1}  \frac{j_l^{\prime}(k \tau_2)}{k \tau_2} \int_0^{\tau_1} d \tau_1^{\prime} \frac{\tau_1^{\prime 8} f(\tau_1^{\prime}, \xi_0)}{k^2 (1-v^2)} B^{V-V}(\tau_1^{\prime}), \\
 \label{Clt2}
&   C_l^{T} = \frac{\kappa^2}{2 \pi} \int_0^{\infty} k^2 dk \frac{(l+2)!}{(l-2)!} \int_0^{\tau0} d \tau_1 \int_0^{\tau0} d \tau_2 \frac{j_l(k \tau_1)}{(k \tau_1)^2}  \frac{j_l(k \tau_2)}{(k \tau_2)^2}   \int_0^{\tau1} d \tau_1^{\prime} k^6 \tau_1^{\prime 8} F(\tau_1^{\prime}, \tau_1) F(\tau_2^{\prime},\tau_2) \frac{f(\tau_1^{\prime},\xi_0)}{k^2(1-v^2)} B^{T-T} (\tau_1^{\prime}),
\end{eqnarray}
where
\begin{eqnarray}
\label{Fsc}
 & F_{sc} =  \frac{1}{9} j_l(k \tau_1) j_l(k \tau_2) \frac{\tau_1^{\prime 4}}{\tau_1^2 \tau_2^2} \biggl( B^{Tr-Tr}(\tau_1^{\prime}) + 4 B^{Tr-S}(\tau_1^{\prime}) + 4 B^{S-S}(\tau_1^{\prime}) \biggl)  + \nonumber \\
& + \frac{1}{3} \left( j_l^{\prime \prime} (k \tau_1) j_l(k \tau_2)  \frac{\tau^{\prime \; 2}}{\tau_2^2} + j_l^{\prime \prime} (k \tau_2) j_l(k \tau_1) \frac{\tau^{\prime \; 2}}{\tau_1^2} \right) \left( B^{Tr-S}(\tau_1^{\prime}) + 2 B^{S-S}(\tau_1^{\prime}) \right) + j_l^{\prime \prime}(k \tau_1) j_l^{\prime \prime} (k \tau_2) B^{S-S}(\tau_1^{\prime}),
\end{eqnarray}
\begin{eqnarray}
\label{Tr-Tr}
\hspace{-0.5in} \text{with trace components: } & \; B^{Tr-Tr}(\tau_1^{\prime}) =\left[1 + v^2 - \tilde{T} / \tilde{U}(1-v^2)  \right]^2 B^{00-00}(\tau_1^{\prime}),\\
\label{Tr-S}
& B^{Tr-S}(\tau_1^{\prime})=\left[1 + v^2 - \tilde{T} / \tilde{U}(1-v^2)  \right] B^{00-S}(\tau_1^{\prime}).
\end{eqnarray}
\end{widetext}

In the final form of equations (\ref{Cls2}) and (\ref{Fsc}) we have expressed the contribution from the ``00" component in terms of the trace component ``$Tr$" using the relations~(\ref{Tr-Tr}) and~(\ref{Tr-S}), which can be derived from~(\ref{Stress-Energy00}) and~(\ref{Stress-EnergyIJ}). It should be stressed that in obtaining equations~(\ref{Cls2}),~(\ref{Clv2}) and~(\ref{Clt2}) we have only used the approximation~(\ref{CorrelatorsForm2}). We have thus succeeded to derive full semi-analytic expressions for the scalar, vector and tensor contributions to the angular powerspectrum from cosmic strings with arbitrary currents, valid in matter domination and under the approximation~(\ref{CorrelatorsForm2}).

In the superhorizon limit $k \tau < 1$ considered above, the two-point correlators have the simple form~(\ref{B001})-(\ref{BT1}) and we can factor out from the integrals~(\ref{Cls2})-(\ref{Clt2}) the key quantities characterising the cosmic string network: $\upsilon$, $\xi_0$, $\hat{U}$, $\hat{T}$ and $Q$. This allows us to establish a direct connection between cosmic string network parameters and the string contribution to CMB anisotropies, valid on superhorizon scales. For the vector~(\ref{Clv2}) and tensor~(\ref{Clt2}) contributions it is easy to see that
\begin{equation}
\begin{gathered}
 \label{FinalClV-ClT}
 \hspace{-0.1in} C_l^{V,T} \sim (G \mu_0)^2 \times \\
 \frac{\hat{U}^2 \upsilon^4 + \hat{T} \hat{U} \upsilon^2 (1-\upsilon^2) +  \hat{T}^2 (1-\upsilon^2)^2 + 3 \upsilon^2 Q^2 }{\xi_0 (1-v^2)}\,,
\end{gathered}
\end{equation}
which agrees with the result of~\cite{ACMS} in the limit $Q=0$, $U=\alpha\mu_0$ and $T=\mu_0/\alpha$. 

The treatment of the scalar mode~(\ref{Cls2}) is more subtle. We will estimate it to leading order, using the following asymptotic form of the spherical Bessel function $j_l(x) \sim x^l$, valid when $0<x<<\sqrt{l+1}$. This approximation is justified when we consider the scalar contribution at large multipole moments $l$. Since the angular power spectrum $C_l$ for cosmic string networks typically peaks at $l>500$ we can take the leading term of~(\ref{Fsc}) as $j_l^{\prime \prime}(k \tau_1) j_l^{\prime \prime}(k \tau_2) B^{S-S}(\tau_1^{\prime})$. It follows that, in this approximation, the scalar contribution will be the same as the above approximate expressions for the vector and tensor components
\begin{equation}
\begin{gathered}
 \label{FinalClS}
 C_{1<<l}^{S} \sim (G \mu_0)^2 \times \\
 \frac{\hat{U}^2 \upsilon^4 + \hat{T} \hat{U} \upsilon^2 (1-\upsilon^2) +  \hat{T}^2 (1-\upsilon^2)^2 + 3 \upsilon^2 Q^2 }{\xi_0 (1-v^2)}.
\end{gathered}
\end{equation}

Let us briefly summarize the results presented in this section. For the action~(\ref{ActionGeneral}) describing a  string with arbitrary current we first derived the stress-energy tensor~(\ref{StressEnergyGeneral}) and obtained the (microscopic) equations of motion~(\ref{EquationMotionGeneral1})-(\ref{EquationMotionGeneral3}), the mass per unit length~(\ref{MassPerUnitLength}) and string tension~(\ref{Tension}). From these, we then developed a macroscopic VOS evolution model~(\ref{AvEquationMotionGeneral1})-(\ref{AvEquationMotionGeneral2}) for a string network with arbitrary currents and used it to estimate analytically the CMB contribution~(\ref{FinalClV-ClT})-(\ref{FinalClS}) from such a string network in the matter domination era. All these results are determined by $\hat{U}$, $\hat{T}$ and $Q$. In particular, changing the form of the function $f(\kappa,\Delta/\gamma)$ in the string action~(\ref{ActionGeneral}) leads to a redefinition of $\hat{U}$, $\hat{T}$ and $Q$ rather than a change in the string equations of motion~(\ref{EquationMotionGeneral1})-(\ref{EquationMotionGeneral2}). 

In the next two sections we will consider two specific physically motivated cases: wiggly and superconducting (chiral) cosmic string networks.

\section{Wiggly model}

In this section we consider the case of wiggly cosmic strings. This model was developed as an effective description of small-scale structure on cosmic strings~\cite{Carter90,Carter95,Vilenkin}. By applying a suitable phenomenological Lagrangian, the evolution of wiggly string networks was studied in~\cite{VieiraMartinsShellard,VieiraMartinsShellard2}. However, a great deal about the dynamics and scaling behaviour of the network can be understood by focusing on the equation of state for wiggly strings, without specifying the precise form of the Lagrangian. 

The equation of state for wiggly strings is
\begin{equation}
\begin{split}
   \label{EqOfStWiggl}
      U T  =& \mu_0^2, \\
    U = \mu_0 \mu,  \; & \; T = \mu_0 / \mu, \\
\end{split}
\end{equation}
or equivalently
\begin{equation}
\begin{split}
   \label{EqOfStWigglhat}
     \hat{U} \hat{T} =& 1\,, \\
    \hat{U} = \mu,  \;  \; \hat{T} = & 1/ \mu, \; \; Q=0\,,
\end{split}
\end{equation}
where $\mu$ is a dimensionless parameter quantifying the amount of wiggles on the string, and $\mu=1$ corresponds to the usual Nambu-Goto string (the same parameter was denoted as $\alpha$ in \cite{PogosianVachaspati}).

Applying the equation of state~(\ref{EqOfStWiggl}) to the averaged equations of motion~(\ref{AvEquationMotionGeneral1}) and~(\ref{AvEquationMotionGeneral2}) we obtain
\begin{eqnarray}
   \label{AverageEquationMotionGeneral1}
&  2 \frac{d L_c}{d \tau} = \frac{\dot{a}}{a} L_c \left[ 1 + \upsilon^{2} - \frac{1-\upsilon^2}{\mu^2} \right], \\
 \label{AverageEquationMotionGeneral2}
& \frac{d\upsilon}{d \tau} = \left( 1 - \upsilon^2 \right) \left[ \frac{k(\upsilon)}{L_c \mu^{5/2}} - \frac{\dot{a}}{a} \upsilon \left( 1 + \frac{1}{\mu^2} \right) \right]\,.
\end{eqnarray}
Note that the comoving correlation length is connected to comoving characteristic length by the following relation $\xi_c = \sqrt{\mu} L_c$. 

We can now include an energy loss term $F(\upsilon, \mu)$ on the right-hand side of equation~(\ref{AverageEquationMotionGeneral1}) and assume scaling behaviour of the network $L_c = \varepsilon \tau$ (while $\xi_c = \xi_0 \tau$). Note that in this case the previously obtained constraint~(\ref{W-restriction}) has the form
\begin{equation}
   \label{Restr}
  v^2 < \frac{2/n - 1 + 1 / \mu^2}{1+1 / \mu^2},
\end{equation}
where, in the scaling regime, $\mu$ is a constant.

\begin{figure}[ht]
\centering
		\includegraphics[width=8.27cm]{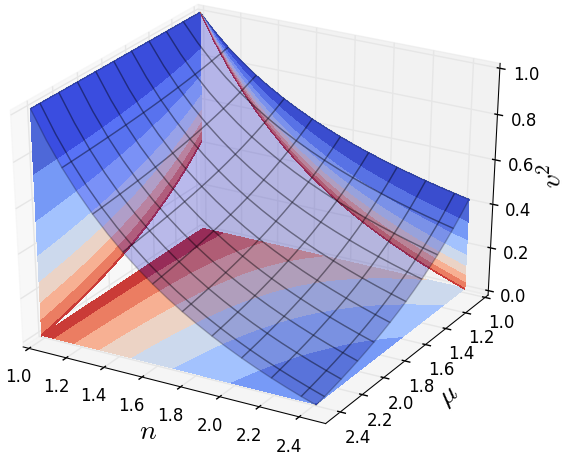}
\caption{The constraint (\protect\ref{Restr}) on the square of the rms velocity, $v^2$, depending on the expansion rate $n$ and the amount of wiggles $\mu$.}
\label{fig:WigConstr}
\end{figure}

The expression~(\ref{Restr}) means that the rms string velocity has an upper limit, determined by the expansion rate $n$ and amount of wiggles $\mu$ on the string; this is illustrated in figure~\ref{fig:WigConstr}. 

It is important to note that this restriction was obtained just by using the equation of state for wiggly cosmic strings~(\ref{EqOfStWiggl}) in our general equation~(\ref{AvEquationMotionGeneral1}). This means that any Lagrangian suitable for wiggly string description (i.e. any choice of $f(\kappa, \Delta/\gamma)$ satisfying~(\ref{EqOfStWiggl}) for the equation of state) cannot change this relation. Moreover, it is valid for any energy loss function $F(v,\mu)$. Thus, any wiggly cosmic string network with any energy loss function of the form $F(v,\mu)$ must satisfy the constraint~~(\ref{Restr}). 

To get a feeling for the size of the maximum network velocity in~(\ref{Restr}) we consider two limiting cases: strings without wiggles ($\mu=1$) and highly wiggly strings ($\mu \rightarrow \infty$):
\begin{eqnarray}
   \label{NoWiggles}
&  v^2 < 1/n \qquad (\mu=1), \\
   \label{ManyWiggles}
&  v^2 < 2/n-1 \qquad (\mu \rightarrow \infty).
\end{eqnarray}

As seen from Eq.~(\ref{NoWiggles}), for strings without wiggles only very fast expansion rates $n$ can cause a significant restriction to the string network velocity, while for highly wiggled strings the limit~(\ref{ManyWiggles}) provides a severe constraint even when $n=2$ (matter domination era). For wiggly strings with $\mu=1.5$ in the matter domination era ($n=2$) the velocity is limited as $v^2<0.3$, which is close to the values of rms velocities from field theory simulations~\cite{HindmarshLizarragaUrrestillaDaverioKunz} in the matter domination era.

Let us now study the full description of the wiggly cosmic string network model~\cite{VieiraMartinsShellard,VieiraMartinsShellard2} described by the action
\begin{equation}
   \label{WigglyAction}
  S = \mu_0 \int \omega \sqrt{-\gamma} d^2 \sigma,
\end{equation}
where $\omega = \sqrt{1-\kappa} $.

 The derivation of the averaged equations of motion for this model can be found in~\cite{VieiraMartinsShellard,VieiraMartinsShellard2}. 
 We will use the final system of equations in the following form (where we have omitted the term responsible for scale dependence)
\begin{eqnarray}
 \label{WigglyEq1}
&  2 \frac{dL_c}{d \tau} = \frac{\dot{a}}{a} L_c \left[ 1 + \upsilon^{2} - \frac{1-\upsilon^2}{\mu^2} \right] + \frac{c f_a \upsilon}{\sqrt{\mu}}, \\
 \label{WigglyEq2}
& \frac{d\upsilon}{d \tau} = \left( 1 - \upsilon^2 \right) \left[ \frac{k}{L_c \mu^{5/2}} - \frac{\dot{a}}{a} \upsilon \left( 1 + \frac{1}{\mu^2} \right) \right], \\
 \label{WigglyEq3}
&  \frac{1}{\mu} \frac{d \mu}{d \tau} = \frac{\upsilon}{L \sqrt{\mu}} \left[ k \left( 1 - \frac{1}{\mu^2} \right) - c (f_a - f_o - S) \right]- \nonumber \\
& \qquad \qquad \qquad -\frac{\dot{a}}{a} \left(1-\frac{1}{\mu^2} \right),
\end{eqnarray}
where the three functions $f_a(\mu)$, $f_0(\mu)$ and $S(\mu)$ quantify energy loss/transfer:
\begin{eqnarray} 
\label{Loop-creation}
& 2 \left( \frac{d \xi}{dt} \right)_{\text{only big loops}} = c f_0(\mu) \upsilon, \\
& 2 \left( \frac{d L}{dt} \right)_{\text{all loops}} = c f_a(\mu) \frac{\upsilon L}{\xi}, \\
& 2 \left( \frac{d \xi}{dt} \right)_{\text{energy transfer}} = c S(\mu) \upsilon\,.
\end{eqnarray}
Here, $c$ is a constant ``loop chopping" parameter (see below), $t=\int a d\tau$, and $\xi=\xi_c a$, $L=L_c a$ are the physical (rather than comoving) lengthscales corresponding to $\xi_c$ and $L_c$.

The term $f_0(\mu)$ accounts for the energy loss due to the formation of big loops. Here, ``big" means that they are formed by intersections of strings separated by distances of order the correlation length $\xi$ or by self-intersections at the scale of the radius of curvature $R \approx \xi$. The function $f_a(\mu)$ describes the energy loss caused by all types of loops, and the difference $f_a(\mu)-f_0(\mu)$ corresponds to the energy loss by small loops only, which is driven by the presence of wiggles. 

In order to reproduce correctly the original model without wiggles, we can use the energy loss/transfer functions as discussed in~\cite{VieiraMartinsShellard2}
\begin{eqnarray} 
\label{energy loss functions}
&  f_0(\mu) = 1, \\
\label{energy loss functions2}
& f_a(\mu) = 1 + \eta \left( 1 - \frac{1}{\sqrt{\mu}} \right), \\
\label{energy loss functions3}
&  S(\mu) = D(1-\frac{1}{\mu^2}),
\end{eqnarray}
where $D$ and $\eta$ are constants.

Thus, the evolution of a wiggly string network is described by the system of ordinary differential equations~(\ref{WigglyEq1})-(\ref{WigglyEq3}), which, in view of equations (\ref{energy loss functions})-(\ref{energy loss functions3}) includes three free constant parameters $c$, $D$ and $\eta$~\cite{VieiraMartinsShellard2}:

$\bullet$ $c$ is the ``loop chopping efficiency" parameter quantifying how much energy the network loses due to the production of ordinary loops;

$\bullet$ $\eta$ is a parameter describing the energy loss enhancement due to the creation of small loops caused by the presence of wiggles; 

$\bullet$ $D$ is a parameter quantifying the amount of energy transferred from large to small scales.

By making various different choices of parameters $c$, $\eta$ and $D$ we can explore the effects of the energy loss/transfer mechanisms described above on the evolution of the string network. Note that $c$ has been measured in Abelian-Higgs and Goto-Nambu simulations to be $c=0.23\pm0.04$ \cite{Moore1,Moore2}, but there are no such measurements for the other two parameters.  Let us study how these phenomenological quantities can change the prediction for the CMB anisotropy caused by wiggly cosmic string networks.

\begin{figure*}[ht]
\centering
		\includegraphics[width=17.5cm]{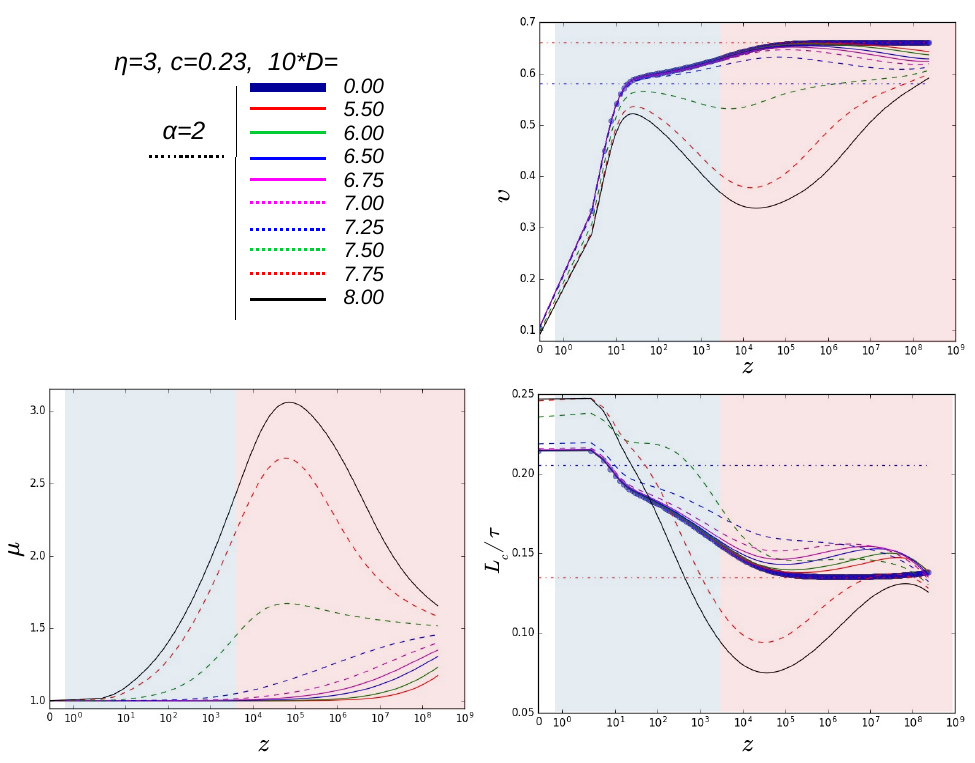}
\caption{Evolution of the rms velocity $\upsilon$, comoving characteristic length $L_c$ and amount of wiggles $\mu$ as a function of redshift $z$ for wiggly cosmic string networks with different values of the parameter $D$, obtained by a modified version of the CMBact code~\cite{PogosianVachaspati}. The horizontal dashed red and blue lines correspond to the usual (without wiggles; $\mu=1$) scaling regimes for radiation (red shaded area) and matter domination (blue shaded area) epochs respectively. Note that the horizontal (redshift) axis is depicted in a linear scale in the redshift range $0<z<1$ and in a logarithmic scale for $z>1$.}
\label{fig:EvolutionWiggly}
\end{figure*}

\begin{figure*}[ht]
\centering
		\includegraphics[width=17.5cm]{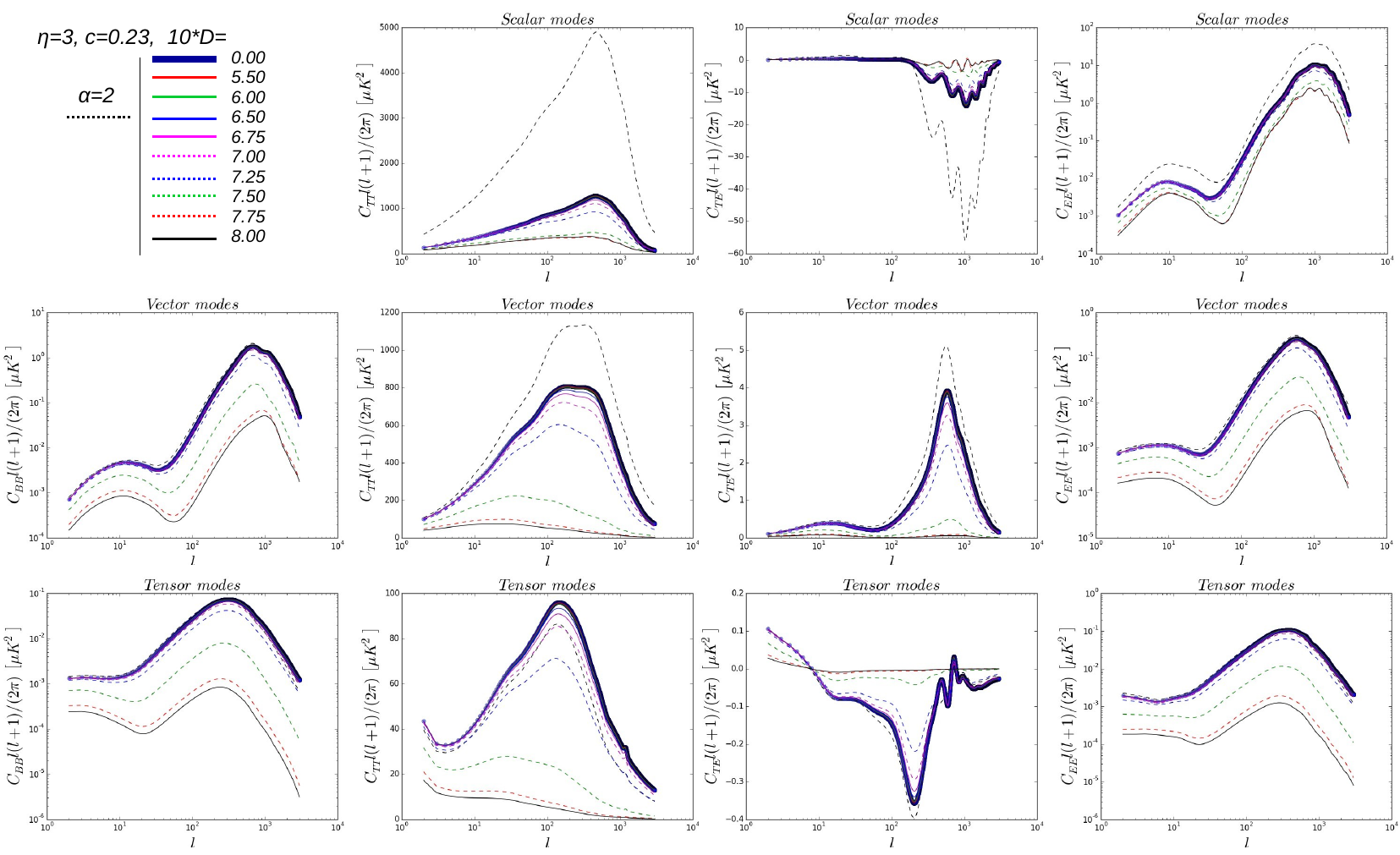}
\caption{CMB anisotropy for wiggly cosmic string networks obtained by a modified version of the CMBact code~\cite{PogosianVachaspati}. The panels show scalar, vector and tensor contributions (top to bottom) of the $BB$, $TT$, $TE$ and $EE$ modes (left to right). (Note there is no $BB$ contribution from scalar modes.) These have been computed for different values of $D$ with fixed $\eta$.  The CMBact result from~\cite{PogosianVachaspati} with $\alpha=2$ is shown by the black dashed line for comparison.}
\label{fig:CMBwiggly}
\end{figure*}

In order to investigate in detail the effects of string wiggles on the predicted CMB anisotropies from cosmic string networks, we implement the wiggly VOS model~(\ref{WigglyEq1})-(\ref{WigglyEq3}) into the CMBact code~\cite{PogosianVachaspati}. The original code was developed so as to take into account the presence of string wiggles in the computation of the string-induced CMB anisotropy. However, in the original CMBact package, wiggles were modelled by a single (constant) phenomenological parameter $\alpha=\mu$ modifying the effective mass per unit length and string tension at the level of the stress-energy tensor~(\ref{StressEnergyWiggly}). In other words, within the approximations of the original CMBact code, the amount of wiggles was not a dynamical parameter and did not influence the equations of motion, while from the wiggly VOS model we have just discussed it is clear that these effects must, in general, be present. Here, we implement the full description of wiggly strings in CMBact. Using the equation of state for wiggly strings~(\ref{EqOfStWiggl}) we first rewrite the stress-energy tensor~(\ref{StressEnergyGeneral}) as 
\begin{equation}
\begin{split}
   \label{StressEnergyWiggly}
    & \qquad \quad T^{\mu \nu} (y) =  \frac{\mu_0}{\sqrt{-g}} \int  d^2 \sigma  \\
    & \left( \epsilon \mu \dot{\textbf{x}}^{\mu} \dot{\textbf{x}}^{\nu}  -  \frac{\textbf{x}^{\prime \mu} \textbf{x}^{\prime \nu}}{\epsilon \mu} \right) \delta^{(4)}(y-x(\sigma))\,,
\end{split}
\end{equation}
where $\mu$ is the amount of wiggles, which is now dynamical, satisfying equation (\ref{WigglyEq3}). The size of string segments is set to be equal\footnote{For an even more realistic model we could consider the strings segments to have a range of sizes and speeds picked from appropriate distributions as in~\cite{Tom}, but here we want to focus on the effects of string wiggles only and compare to the results of the original CMBact code, which also takes all segments to have the same size and speed.} to the correlation length $\xi_0 \tau$. We also change the VOS equations of motion in CMBact to the full system~(\ref{WigglyEq1})-(\ref{WigglyEq3}) and implement the stress-energy components~(\ref{Stress-EnergyChSc})-(\ref{Stress-EnergyChTens}). With these modifications, we achieve a full treatment of wiggly cosmic string networks in CMBact.

In figure~\ref{fig:EvolutionWiggly} we show our results for network evolution and in figure~\ref{fig:CMBwiggly} the corresponding CMB anisotropies computed in our modified version of CMBact. In both figures we also show the corresponding results of the original CMBact code~\cite{PogosianVachaspati} for comparison. Regarding figure~\ref{fig:EvolutionWiggly}, we note that the accuracy of CMBact is comparatively worse at low redshifts; this explains why the effects of the matter to acceleration transition seemingly become visible around redshifts of a few, while the onset of acceleration occurs below $z=1$. This point is not crucial for our analysis, since our goal is to make a comparative study of the effects of the additional degrees of freedom on the strings. Moreover, these low redshifts have a relatively small effect on the overall CMB signal. Nevertheless, this is an issue which should be addressed if this code is to be used for quantitative comparisons with current or forthcoming CMB data.
 
We have chosen to vary parameter $D$, keeping $\eta$ fixed, which allows us to cover a wide range of $\mu$ values. Fixing $D$ and increasing $\eta$ is equivalent to decreasing the amount of wiggles and an effective change of $c$, which is already covered from our variation of $D$ with fixed $\eta$. It is also important to note that in order to have an attractor scaling solution when $a \propto \tau^n$ the following condition must be satisfied
\begin{equation} 
\label{etaDCond}
  \eta > \frac{D\left( 1 - 1/\mu^2 \right) }{1-1/\sqrt{\mu}}.
\end{equation}

Physically, this means that in order to achieve a scaling solution, small scale structure should be able to lose energy (controlled by parameter $\eta$) faster than it receives the energy from large scales (controlled by parameter $D$). When the condition~(\ref{etaDCond}) is violated, energy accumulates at small scales and there is no stable scaling regime for these wiggly cosmic strings. In practice, the condition~(\ref{etaDCond}) is used as a guide for estimating the range of variation of $D$. 

Figure~\ref{fig:CMBwiggly} shows how the full treatment of wiggly cosmic string networks affects the prediction for the string-induced CMB anisotropy. Note that the CMB contribution is generally smaller than for ordinary cosmic strings (i.e. without wiggles, $\mu=1$). This is mainly due to a reduction in the rms string velocity $\upsilon$ (see figure~\ref{fig:EvolutionWiggly}) when the amount of wiggles $\mu$ increases. In view of the observed changes to the usual CMB predictions for cosmic strings, we argue that to achieve accurate results for wiggly cosmic strings, one should study them in the framework of the complete wiggly model~(\ref{WigglyEq1})-(\ref{WigglyEq3}) and the modified version of CMBact developed here. This generally leads to a weakening of the CMB-derived constraint on the string tension $\mu_0$ (but note that there is also a region in parameter space -- for large $D$ -- where the correlation length can actually become smaller than for ordinary strings, see figure \ref{fig:EvolutionWiggly}). 

Note that both the evolution and CMB results from our wiggly VOS model are somewhat closer in comparison to results from Abelian-Higgs simulations (and similarly ordinary VOS results are closer to Nambu-Goto simulations). It is then tempting to speculate that wiggles play a dynamical role analogous to that of the averaged field fluctuations that appear in Abelian-Higgs field theory simulations (as opposed to effective Nambu-Goto simulations). This hypothesis may be investigated by direct comparisons of Abelian-Higgs and Goto-Nambu simulations with suitably high resolutions and dynamic ranges.

To end this section, let us return to the wiggly model but this time without referring to the specific Lagrangian~(\ref{WigglyAction}). We wish to study the scaling regime for wiggly strings but leaving the amount of wiggles $\mu$ as a free parameter that we can tune. Instead of varying parameter $D$, as it was done above, we can vary $\mu$. This approach does not require an assumption on the energy transfer function~(\ref{energy loss functions3}); we only need to define how energy loss depends on the amount of wiggles~(\ref{energy loss functions2}). Let us now estimate how the rms velocity $\upsilon$ and comoving characteristic length $L_c$ are related to the parameters $c$, $\eta$ and $\mu$ in the scaling regime. We insert the scaling solution $L_c = \varepsilon \tau$, $\upsilon=$const to equations~(\ref{AverageEquationMotionGeneral1}), (\ref{AverageEquationMotionGeneral2}) to obtain the algebraic equations
\begin{eqnarray}
 \label{WigglyEqScale1}
&  \varepsilon \left(2-n \left[ 1 + \upsilon^{2} - \frac{1-\upsilon^2}{\mu^2} \right] \right) =  \frac{ c f_a(\mu) \upsilon}{\sqrt{\mu}} , \\
 \label{WigglyEqScale2}
&   \frac{k(\upsilon)}{\varepsilon \mu^{5/2}} = n \upsilon \left( 1 + \frac{1}{\mu^2} \right),
\end{eqnarray}
where we have included energy loss function $f_a(\mu)$ given by~(\ref{energy loss functions2}).

\begin{figure*}[ht]
\centering
		\includegraphics[width=8.5cm]{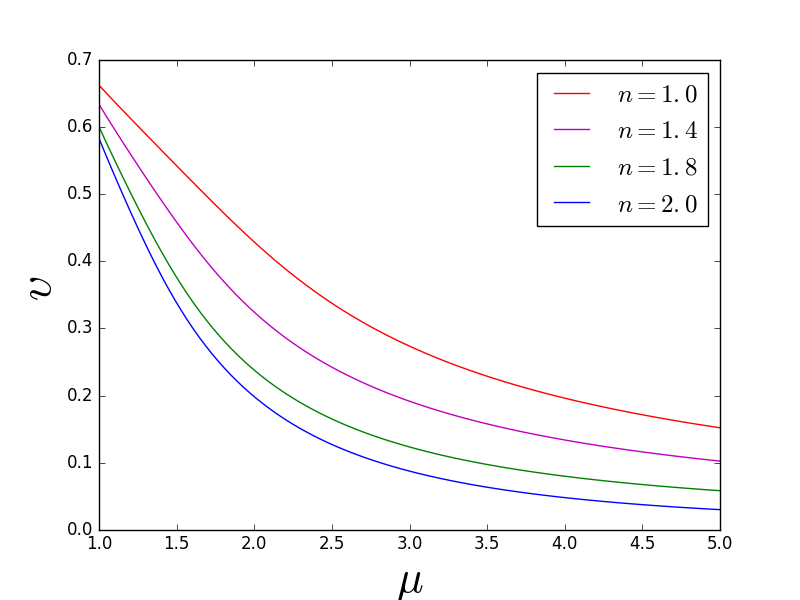}
		\includegraphics[width=8.5cm]{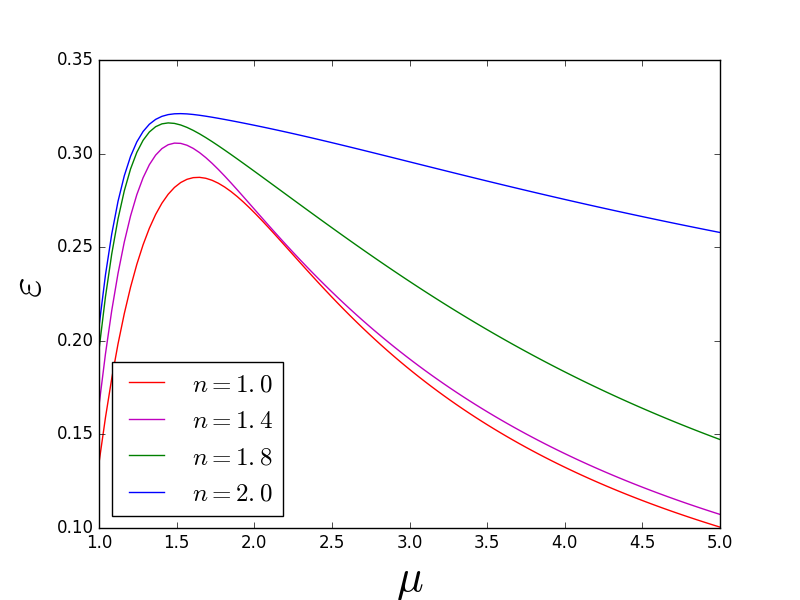} 

\caption{Dependence of the scaling values of the rms velocity, $v$, and the comoving correlation length divided by conformal time, $\varepsilon$, on the amount of wiggles $\mu$ for different expansion rates $n$. }
\label{fig:VelCorMu}
\end{figure*}

\begin{figure}[ht]
\centering
		\includegraphics[width=8.0cm]{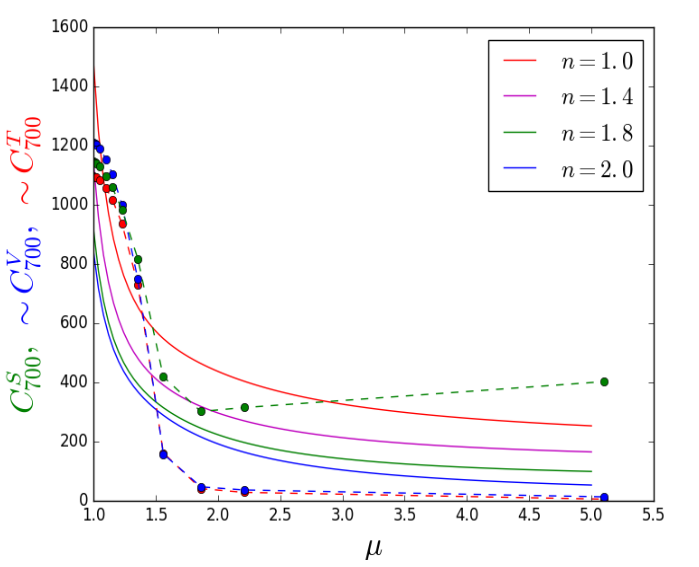}		
\caption{Comparison between the behaviour of the string-induced angular power spectrum $C_l$ for different amount of wiggles in our analytic approximation (solid lines) and the numerical computation using our modified CMBact code (circles). The dependence on $\mu$ has been estimated analytically using equations~(\ref{FinalClS}), (\ref{FinalClV-ClT}) together with the equations for the scaling regime of the network~(\ref{WigglyEqScale1})-(\ref{WigglyEqScale2}). Using the value of $\mu$ in the matter domination era and $C_l$'s for scalar (green), vector (blue) and tensor (red) components at $l=700$ (where the sum peaks), we have obtained the $C_l - \mu$ dependence from the CMBact code.}
\label{figure5}
\end{figure}

Despite the reduction of the equations of motion to algebraic equations~(\ref{WigglyEqScale1}) and~(\ref{WigglyEqScale2}) in the scaling regime, it is still not possible to solve them analytically, mainly due to the complicated form of the momentum parameter~(\ref{MomentumParameter}). To study how the amount of wiggles affects the macroscopic parameters $\upsilon$ (rms string velocities) and $\varepsilon$ (comoving correlation length in units of conformal time) in the scaling regime we solve the system~(\ref{WigglyEqScale1})-(\ref{WigglyEqScale2}) numerically for different expansion rates $n$. The results are shown in figure~\ref{fig:VelCorMu}. It is seen that the rms velocity $\upsilon$, as anticipated from the restriction~(\ref{Restr}), decreases with the growth of the amount of wiggles $\mu$. This is also in agreement with our results for the rms velocity evolution (see figure~\ref{fig:EvolutionWiggly}) in the dynamical wiggly model for a realistic expansion history. The situation for $\varepsilon$ is more interesting. The correlation length does not increase monotonically with the amount of wiggles but has a maximum around $\mu=1.5-1.9$. This is also in agreement to our full treatment in figure~\ref{fig:EvolutionWiggly} where we modelled string wiggles by varying parameter $D$ and took a realistic expansion history.

Since we have computed the velocity $\upsilon$ and correlation length $\xi_0 \tau = \sqrt{\mu} \varepsilon \tau$ in the scaling regime, we can use equations~(\ref{FinalClV-ClT}) and~(\ref{FinalClS}) to estimate how the contribution to the CMB anisotropy from cosmic strings depends on the amount of string wiggles. For wiggly cosmic strings the angular power spectrum $C_l$ has the following dependence (which coincides with the result in~\cite{ACMS})
\begin{equation}
 \label{Clwiggly}
 C_l \sim (G \mu_0)^2 \frac{\mu^4 \upsilon^4 + \mu^2 \upsilon^2 (1-\upsilon^2)+  (1-\upsilon^2)^2 }{\mu^2 \xi_0 (1-v^2)},
\end{equation}
where scalar, vector and tensor components depend on string parameters in the same way.

We can now compare the dependence in equation~(\ref{Clwiggly}) with our numerical results using our modified CMBact code. By choosing the $\mu$ value for the matter domination era and looking at the peak ($l \approx 700$) of the sum of the scalar, vector and tensor contributions we plot them in comparison to the analytic estimate from~(\ref{FinalClV-ClT}) and~(\ref{FinalClS}). This comparison is shown in figure~\ref{figure5}. For our approximate estimate it is seen that after a fast decrease of $C_l$'s with growing amount of wiggles $\mu$, the value of $C_l$ reaches a plateau. A similar behaviour is seen for vector, tensor and scalar components obtained from the full treatment using our modified CMBact code, even though the agreement is somewhat weaker for the scalar contribution. These results reaffirm the approximations used to estimate the analytic dependence of $C_l$ on the string network characteristics.

\section{Superconducting model \newline (chiral case)}

Another special case of current-carrying cosmic strings of notable physical interest is the case of superconducting cosmic strings. This type of strings has been studied thoroughly in the framework of field-theory~\cite{Witten,Ringeval1,LilleyPeteMartin,Ringeval2,LilleyMarcoMartinPeter,HillWdrow,HindmarshCurrent}. In all these cases the stress-energy tensor on the string worldsheet has the following form:
\begin{eqnarray}
\label{FieldStressEnergy}
    T^{a}_b = \begin{pmatrix} A + B & -C \\ C & A - B \end{pmatrix}\,.
\end{eqnarray}
where $A$ arises from the field responsible for the string core formation, while $B$ and $C$ represent additional contributions due to coupling with external fields (dynamics of currents). The stress-energy tensor~(\ref{FieldStressEnergy}) is written for the worldsheet metric $\eta^{ab} = \begin{pmatrix}  1 & 0 \\ 0 & -1 \end{pmatrix}$ on a 4-dimensional Minskowski spacetime background with $\epsilon = 1$.

Consider now the two-dimensional stress-energy tensor for the action~(\ref{ActionGeneral}), which reads
\begin{eqnarray}
\label{ChiralStressEnergy}
    T^{a}_b = \begin{pmatrix}  \mu_0 \tilde{U} & - \mu_0 \frac{\Phi}{\epsilon} \\  \mu_0 \epsilon \Phi &  \mu_0 \tilde{T} \end{pmatrix}.
\end{eqnarray}

There is an obvious correspondence between the stress-energy tensors~(\ref{FieldStressEnergy}) and~(\ref{ChiralStressEnergy}); they are in agreement if we demand the chiral condition~\cite{Vortons,OliveiraAvgoustidisMartins}
\begin{equation}
\label{ChiralCond}
    \kappa \rightarrow 0\,,
\end{equation}
which also means
\begin{equation}
\label{ChiralCondd}
    \Delta \rightarrow 0\,;
\end{equation}
here $\kappa$ and $\Delta$ are defined by Eq. (\ref{Terms}). These imply that $\tilde{U} = 1+\Phi$, $\tilde{T} = 1 - \Phi$. In Minkowski space ($\epsilon=1$) we see that $A=\mu_0$, $B = \mu_0 \Phi$ and $C=\mu_0 \Phi$, so we have the condition $B=C$. In order to avoid this situation and be able to reproduce a stress-energy tensor of the form (\ref{FieldStressEnergy}) within the Nambu-Goto approximation, we need to use at least two scalar fields. It has already been demonstrated that adding any number of additional fields~(\ref{kappa}) together with the definitions~(\ref{U2})-(\ref{Phi2}) keeps the evolution equations~(\ref{EquationMotionGeneral1})-(\ref{EquationMotionGeneral2}) unchanged, replacing the scalar field equation~(\ref{EquationMotionGeneral3}) by the set of equations~(\ref{EquationMotionGeneral32}). In effect, introducing additional fields makes $C$ and $B$ different in Minkowski space.

Indeed, when we add extra scalar fields we obtain a stress-energy tensor in the form of (\ref{FieldStressEnergy}) with the correspondence\footnote{Here we used an assumption that all multipliers $\frac{\partial f}{\partial \Delta_i}$ are equal as well as all $\frac{\partial f}{\partial \kappa_i}$ are equal (\ref{ChiralCond}). } $A = \mu_0$, $B=2 \mu_0 \gamma^{00} \left( \frac{\partial f}{\partial \kappa} - \frac{\partial f}{\partial \Delta} \right) \sum_i  \dot{\varphi_i}^2 = \mu_0 \Psi$ and $C = \mu_0 \Phi$, where $\Phi$ is given by equation (\ref{Phi2}) and we have assumed a Minkowski background. These correspond to
\begin{equation}
\begin{split}
   \label{EqOfStCh}
   \hat{U} = 1+\left<\Psi\right>,  \; & \; \hat{T} = 1-\left<\Psi\right>, \; \; Q=\left<\Phi\right>.
\end{split}
\end{equation}
Thus, this multiple worldsheet field approach provides enough flexibility to reproduce the field-theoretical stress-energy tensor variables in~(\ref{FieldStressEnergy}) within the Nambu-Goto approximation.

Let us now consider the equations of motion for chiral currents. We will apply our averaging procedure to the system of equations~(\ref{EquationMotionGeneral32}) for the currents, similarly to what we already did for first two equations~(\ref{AvEquationMotionGeneral1}) and~(\ref{AvEquationMotionGeneral2}) for the correlation length and string velocity. First of all, we note that in order to have the appropriate Nambu-Goto limit for the action~(\ref{ActionGeneral}) when $\varphi=0$ we need to have $f(\kappa, \Delta)\xrightarrow[\kappa\to0]{}1$ and additionally $\frac{\partial f(\kappa, \Delta)}{\partial \kappa} \xrightarrow[\kappa\to0]{}  \text{const}$ (as well as $\frac{\partial f(\kappa, \Delta)}{\partial \Delta} \xrightarrow[\Delta\to0]{}  \text{const}$). These conditions allow us to make simplifications, similar to what was done in reference~\cite{OliveiraAvgoustidisMartins}, and consider the case of conserved microscopic charges for each field
\begin{eqnarray}
   \label{ChargeCond1}
    & \epsilon \dot{\varphi_i}= \phi_{i}=\text{const} \;, \\
    \label{ChargeCond2}
    & \varphi_i^{\prime}= \psi_{i}=\text{const} \; ,
\end{eqnarray}
which leads to the additional condition $\epsilon^{\prime}=0$. Furthermore, in this case we can define $\Psi$ and $\Phi$ as

\begin{equation}
   \label{PsiSolution}
     \Psi =  \left( \frac{\partial f}{\partial \kappa} + \frac{\partial f}{\partial \Delta} \right) \sum_i \frac{\phi_i^2}{a^2 \textbf{x}^{\prime  2}},
\end{equation}
\begin{equation}
   \label{PhiSolution}
     \Phi =  \left( \frac{\partial f}{\partial \kappa} + \frac{\partial f}{\partial \Delta} \right) \sum_i \frac{\phi_i \psi_i}{a^2 \textbf{x}^{\prime  2}},
\end{equation}
and~(\ref{ChiralCond}) gives us 
\begin{equation}
   \label{ChiralCond2}
     \sum_i \phi_i^2  = \sum_i \psi_i^2 .
\end{equation}

Expressions~(\ref{PsiSolution}) and~(\ref{PhiSolution}) tell us that if we use the condition of conserved microscopic charges~(\ref{ChargeCond1})-(\ref{ChargeCond2}) we have two variables $\Psi$ and $\Phi$ which evolve in the same way and differ only by a multiplicative constant $\beta$:
\begin{equation}
   \label{AlphaConst}
    \Psi \beta = \Phi ,
\end{equation}
where $\beta = \frac{\sum_i \phi_i \psi_i  }{\sum_i \phi_i^2  }$. Together with~(\ref{ChiralCond2}), this implies that $0<\beta<1$. 

By direct differentiation of equation~(\ref{PsiSolution}) we obtain the following evolution equations for the field $\Psi$ (clearly, the same equations are also obeyed by $\Phi$)
\begin{eqnarray}
   \label{PhiDot}
    & \dot{\Psi} + 2 \frac{\dot{a}}{a} \Psi = 2 \Psi \frac{\dot{\textbf{x}} \cdot \textbf{x}^{\prime \prime}}{\textbf{x}^{\prime \, 2}}, \\
    \label{PhiPrime}
    & \Psi^{\prime} + 2 \Psi  \frac{\textbf{x}^{\prime} \cdot \textbf{x}^{\prime \prime}}{\textbf{x}^{\prime \, 2}} = 0.
\end{eqnarray}

Following the approach of~\cite{OliveiraAvgoustidisMartins}, we average the equations of motion~(\ref{PhiDot}),~(\ref{PhiPrime}) and substitute the equation of state~(\ref{EqOfStCh}) into equations~(\ref{AvEquationMotionGeneral1}) and~(\ref{AvEquationMotionGeneral2}). This leads to the VOS model for superconducting chiral strings, taking into account energy and charge losses (for details on these loss terms see~\cite{OliveiraAvgoustidisMartins})
\begin{eqnarray}
 \label{ChEq1}
&  \frac{dL_c}{d \tau} = \frac{\dot{a}}{a} L_c \frac{\upsilon^{2} + Q}{1+Q} + \upsilon \left( \frac{Q s \beta }{(1+Q)^{3/2}} + \frac{c}{2} \right), \\
 \label{ChEq2}
& \hspace{-0.1in} \frac{d\upsilon}{d \tau} = \frac{1-\upsilon^2}{1+Q}\left[ \frac{k(\upsilon) }{L_c \sqrt{1+Q}} \left( 1 - Q (1+\frac{2 s  \beta}{k(\upsilon)}) \right) - 2 \frac{\dot{a}}{a} \upsilon   \right], \\
 \label{ChEq3}
&  \hspace{-0.1in} \frac{d Q}{d \tau} = 2 Q \left( \frac{k(\upsilon) \upsilon}{L_c \sqrt{1+Q}} -  \frac{\dot{a}}{a}\right) + \frac{c \upsilon \left( 1- \sqrt{1+Q} \right)}{L_c } \sqrt{1+Q} \,.
\end{eqnarray}
We have used the assumption $\left\langle \frac{\Psi^{\prime}}{\epsilon (1+\Psi)} \right\rangle = -s \frac{\upsilon}{R_c} \frac{2 Q}{1+Q}$ ~\cite{OliveiraAvgoustidisMartins} and that the correlation and characteristic lengths are related by $\xi_c=L_c \sqrt{1+Q}$. 

Therefore, our general analysis of chiral current dependence in the action~(\ref{ActionGeneral}) including the addition of extra worldsheet fields has not introduced significant changes in the macroscopic equations describing superconducting chiral cosmic string networks, as compared to the results in~\cite{OliveiraAvgoustidisMartins} (the only difference is that the constant $s$ has now been changed to $\beta s$). Note also that the final result does not depend explicitly on the precise form of the Lagrangian; the important physics can be encoded in the equations of state of the strings, in agreement with our previous discussion. 

The evolution of string networks described by equations~(\ref{ChEq1})-(\ref{ChEq3}) was carefully studied in~\cite{OliveiraAvgoustidisMartins}. It was shown that these networks have generalized scaling solutions\footnote{In this context, by ``generalized scaling solutions" we mean that all three quantities $L_c/\tau$, $v$ and $Q$ approach constant non-zero values (and so the strings have non-zero charge). For larger expansion rates there are also solutions with a decaying charge $Q$ for which $L_c/\tau$ and $v$ are (non-zero) constants but $Q$ approaches zero in a power-law fashion~\cite{OliveiraAvgoustidisMartins}. These correspond to the standard linear scaling solutions of (uncharged) Nambu-Goto strings and we do not discuss them here in detail.} only if the following relation is satisfied
\begin{equation}
   \label{Qscaling}
     n=\frac{2 k(\upsilon)-c \tilde{W}}{c+k(\upsilon)},
\end{equation}
where $\tilde{W}=\frac{\sqrt{1+Q_s}-1}{1+Q_s^{-1}}$, with $Q_s$ a constant corresponding to the scaling value of the function $Q$. As we can see from equation~(\ref{Qscaling}), the expansion rate $n$ for scaling solutions (with constant charge) cannot be larger than $n \leq 2$. The maximal value of $n$ is reached when $c=0$, while for $c=0.23$ (which we use here) we have the condition $n \lesssim 1.6$ for scaling behaviour. For expansion rates $n$ larger than the right hand side of (\ref{Qscaling}) the charge $Q$ on the string decays.

\begin{figure}[ht]
\centering
		\includegraphics[width=8.27cm]{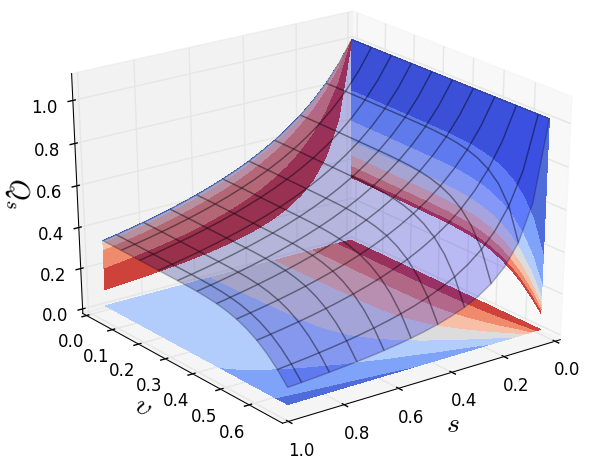}
\caption{Constraints on the possible values of the charge $Q_s$ depending on the rms velocity $\upsilon$ and parameter $s$.}
\label{fig:QRange}
\end{figure}

It is worth noting that at the same time the asymptotic value of the charge $Q_s$ is limited from equation~(\ref{ChEq2}) to satisfy the following (see figure~\ref{fig:QRange})
\begin{equation}
   \label{Qrange}
     Q_s < \frac{k(\upsilon)}{2 s \beta + k(\upsilon)}.
\end{equation}

\begin{figure*}[ht]
\centering
		\includegraphics[width=17.5cm]{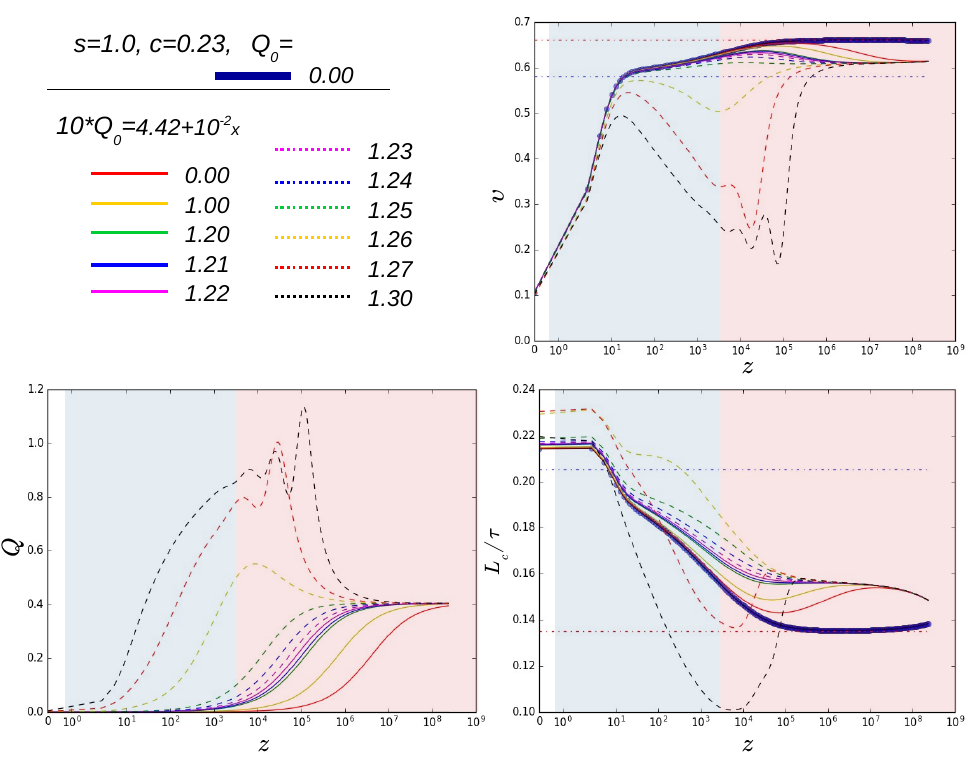}
\caption{Evolution of the rms velocity $\upsilon$, comoving characteristic length $L_c$ and charge $Q$ depending on redshift $z$ for superconducting (chiral) cosmic string networks with different initial conditions $Q_0$ for the string charge, obtained by a modified version of the CMBact code~\cite{PogosianVachaspati}. The horizontal dashed red and blue lines correspond to the usual (without charge, $Q=0$) scaling regimes for radiation (red shaded area) and matter domination (blue shaded area) eras respectively. Note that the horizontal (redshift) axis is depicted in a linear scale in the redshift range $0<z<1$ and in a logarithmic scale for $z>1$.}
\label{fig:EvolutionSuper}
\end{figure*}

\begin{figure*}[ht]
\centering
		\includegraphics[width=17.5cm]{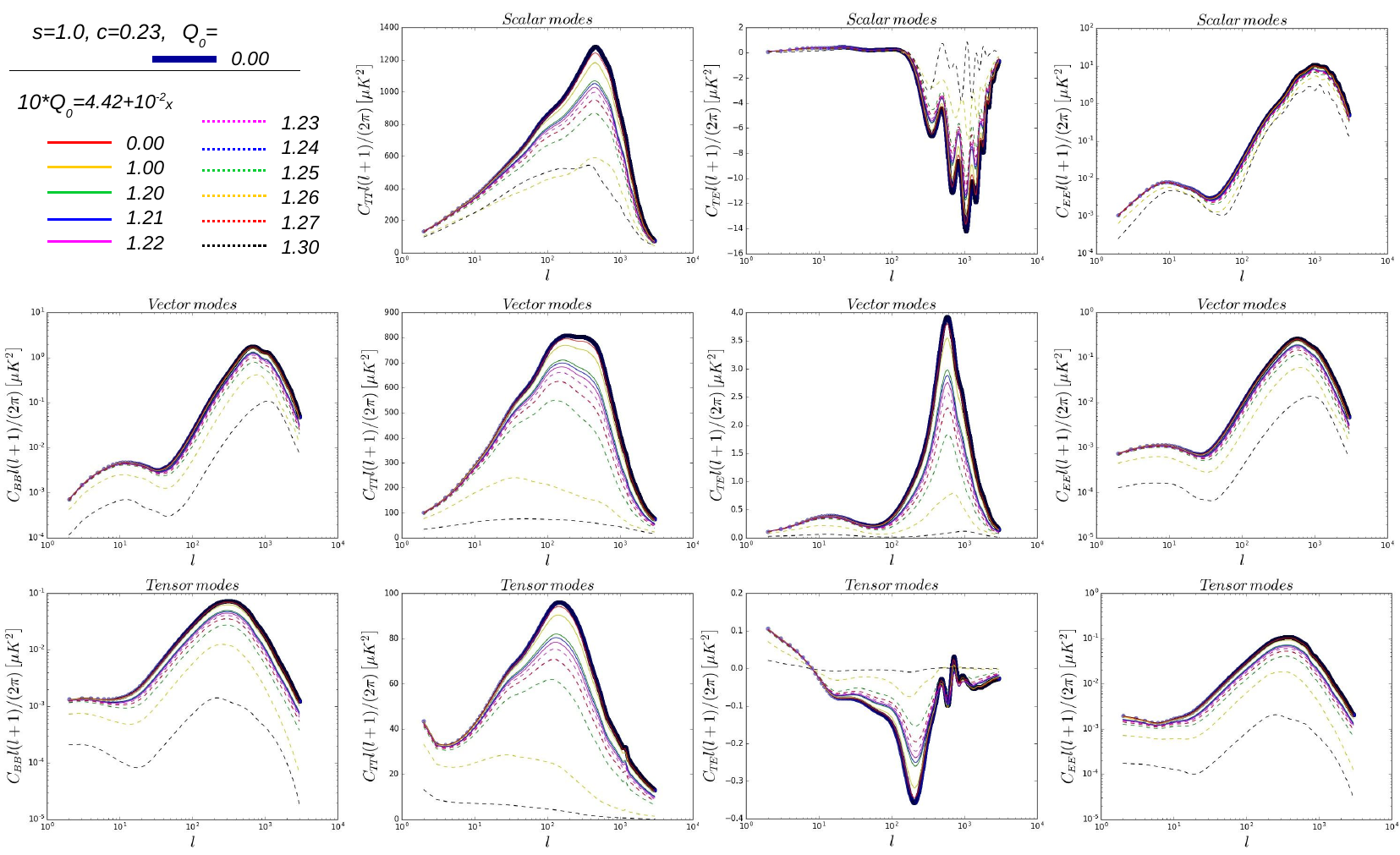}
\caption{CMB anisotropy results for superconducting (chiral) cosmic sting networks obtained by our modified version of CMBact~\cite{PogosianVachaspati}. The panels show the scalar, vector and tensor contributions (top to bottom) to the $BB$, $TT$, $TE$ and $EE$ power spectra (left to right). The calculations are done for different initial conditions of the charge $Q_0$.}
\label{fig:CMBsuper}
\end{figure*}

For all other expansion rates that do not satisfy conditions~(\ref{Qscaling}) and~(\ref{Qrange}), there is no scaling regime with nonzero $Q$. However, in all cases (even in the absence of scaling solutions) we can still evolve the network with our modified VOS model and use equations~(\ref{Stress-EnergyChSc})-(\ref{Stress-EnergyChTens}) with the stress-energy tensor
\begin{eqnarray}
   \label{StressEnergyChiral}
    & T^{\mu \nu} (y) =  \frac{\mu_0}{\sqrt{-g}} \int \sqrt{-\gamma} \biggl( (1+\Psi) u^{\mu} u^{\nu} - (1-\Psi) v^{\mu} v^{\nu} - \nonumber \\
    &   - \alpha \Psi (u^{\mu} v^{\nu} + v^{\mu} u^{\nu}) \biggl)\delta^{(4)}(y-x(\sigma)) d^2 \sigma
\end{eqnarray}
to modify the CMBact code for a superconducting chiral cosmic string network. In the absence of scaling the charge $Q$ for the cosmic string network evolution is controlled mainly by the initial condition $Q_0$. Note that, unlike the wiggly case, there are currently no numerical simulations which can provide us with benchmarks for the value of this charge. Thus, by varying $Q_0$ we obtain different evolutions for cosmic superconducting string networks (see figure~\ref{fig:EvolutionSuper}) and their corresponding contributions to the CMB anisotropy (see figure~\ref{fig:CMBsuper}). 
\begin{figure*}[ht]
\centering
		\includegraphics[width=8.5cm]{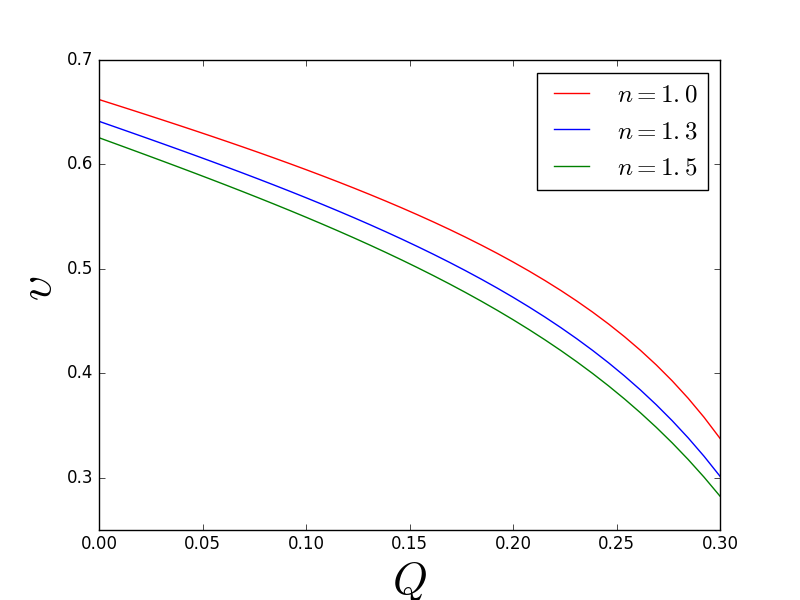}
		\includegraphics[width=8.5cm]{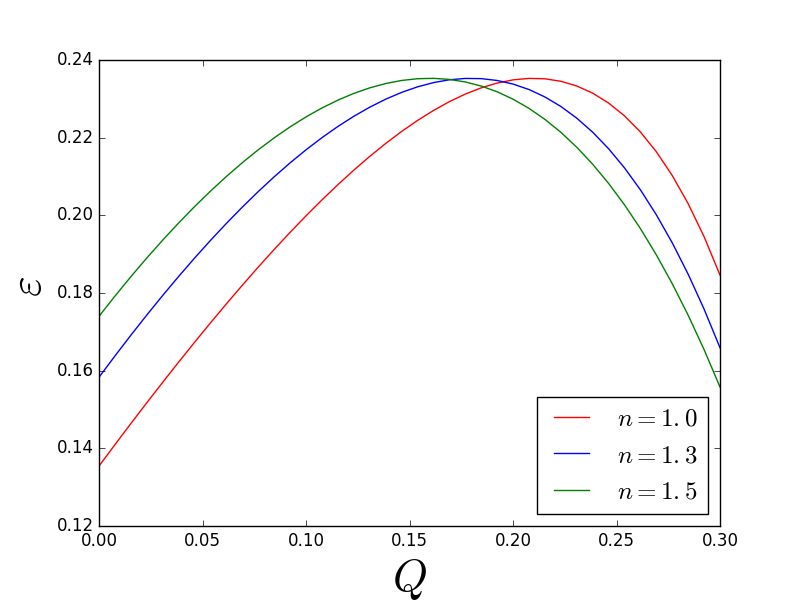} 
\caption{Scaling values of rms velocity, $v$, and comoving correlation length divided by conformal time, $\varepsilon$, depending on the charge $Q$, for different expansion rates $n$. }
\label{fig:VelCorQ}
\end{figure*}
\begin{figure}[ht]
\centering
		\includegraphics[width=8.5cm]{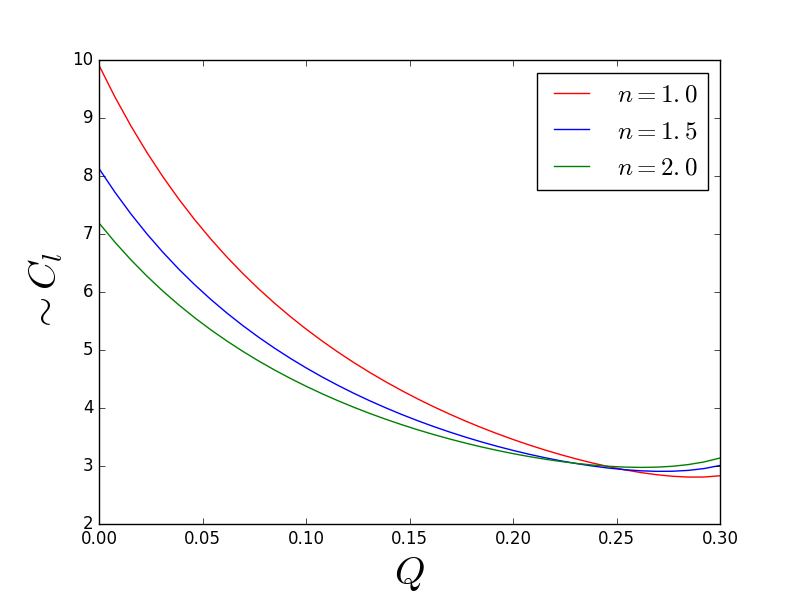}		
\caption{Behaviour of $C_l$ for different string charge $Q$, obtained from the analytical approximation.}
\label{fig:ClQ}
\end{figure}

Let us consider the network at specific values $n$ satisfying equations~(\ref{Qscaling}),~(\ref{Qrange}). For that we will use the typical scaling ansatz $L_c = \varepsilon \tau$ with constant $\varepsilon$, $\upsilon$ and $Q$ in equations~(\ref{ChEq1})-(\ref{ChEq2})
\begin{eqnarray}
 \label{ChEqAlg1}
&  \varepsilon = n \varepsilon \frac{v^{2} + Q}{1+Q} + v \left( \frac{Q s \alpha }{(1+Q)^{3/2}} + \frac{c}{2} \right), \\
 \label{ChEqAlg2}
& \frac{k(v) }{\varepsilon \sqrt{1+Q}} \left( 1 - Q (1+\frac{2 s \alpha}{k(v)}) \right) = 2 n \upsilon.
\end{eqnarray}

Equations~(\ref{ChEqAlg1}) and~(\ref{ChEqAlg2}) describe the network evolution in the scaling regime (they will be valid only in a range of $n$ satisfying (\ref{Qscaling})). Using these equations and the analytic form of the angular power spectrum's dependence on string network parameters~(\ref{FinalClS}), we can make an estimation of $C_l$ for the network of superconducting chiral strings
\begin{equation}
\begin{gathered}
 \label{SuperClS}
 C_l \sim (G \mu_0)^2 \frac{\upsilon^4 (1+3 \beta^2 Q^2)}{\varepsilon (1-v^2)} + \\
\frac{v^2(3 Q^2 (1-\beta^2) + 4 \beta Q -1)+(1-\beta Q)^2}{\varepsilon (1-v^2)},\\
\end{gathered}
\end{equation}
where scalar, vector and tensor components depend on string parameters in the same way.

We can then solve the algebraic equations~(\ref{ChEqAlg1}),~(\ref{ChEqAlg2}) numerically for different values of $Q$ (some solutions are shown in figure~\ref{fig:VelCorQ}) and insert them in~(\ref{SuperClS}) to get an estimate of how the angular power spectrum $C_l$ depends on the value of the charge $Q$ (figure~\ref{fig:ClQ}).

In both our numerical calculations using the modified CMBact code (figures~\ref{fig:EvolutionSuper},~\ref{fig:CMBsuper}) and in our analytic estimates (figures~\ref{fig:VelCorQ},~\ref{fig:ClQ}) we observe that the string rms velocity tends to decrease as we increase the charge $Q$. The comoving correlation length in units of the conformal time $\varepsilon$ increases for small values of the charge, but then reaches a maximum and eventually decreases for higher values of the charge $Q$. Concerning the angular power spectra $C_l$, it should be noted that it is difficult to make extensive comparisons in the case of superconducting strings, as there is no scaling behaviour in the full range of expansion rates $n$ and we do not know which $Q$ values we should choose from our numerical results in figure~\ref{fig:CMBsuper}. However, it is clear that the analytic approach and numerical computation are in qualitative agreement. In particular the angular power spectrum $C_l$ decreases as we increase the charge $Q$ on the string.

\section{Conclusions}

There are many well-motivated scenarios in early universe physics that can leave behind relic defects in the form of cosmic strings. These relics can be utilised as ``fossils" for cosmological research, helping us to obtain a better understanding of the physical processes that took place in the early universe. By developing an accurate description of the evolution of cosmic string networks and using it to calculate quantitative predictions of string-induced observational signals, we can obtain strong constraints on theoretical models leading to a better understanding of early universe physics. Here, we presented a detailed study of the evolution of cosmic strings with currents and demonstrated how the presence of worldsheet currents affects the predictions for the CMB anisotropy produced by cosmic string networks.

In section II we considered the action~(\ref{ActionGeneral}) describing strings with an arbitrary dependence on worldsheet currents. We have described how to average the microscopic equations of motion for this model to obtain macroscopic evolution equations (without energy loss) for the string network~(\ref{AverageEquationMotionGeneral1})-(\ref{AverageEquationMotionGeneral2}). These describe the time evolution of the rms string velocity $\upsilon$ and characteristic length $L$, and depend only on three parameters $\hat{U}$, $\hat{T}$ and $Q$ defining the string equation of state. These same parameters, together with the network quantities $L$ and $\upsilon$, appear directly in the string stress-energy tensor~(\ref{StressEnergyGeneral}) which seeds the string-induced CMB anisotropy. This provides a direct connection between modelling string evolution and computing CMB anisotropies from cosmic string networks, which has allowed us to obtain simple analytic estimates for the dependence of the string angular power spectrum $C_l$ on macroscopic network parameters~(\ref{FinalClV-ClT})-(\ref{FinalClS}). For a more complete semi-analytic treatment of the CMB anisotropy for strings with currents, we have adapted the methodology of \cite{Tom} and have provided coefficients for the relevant integrals in the Appendix. In sections III and IV we considered two specific cases of strings with currents: wiggly and superconducting cosmic strings respectively. In each case we computed the CMB signal numerically using appropriately modified versions of CMBact.    

For wiggly string networks (section III) we studied the specific case when the parameter $\kappa$ in~(\ref{Terms}) only carries a time dependence, $\kappa=\kappa(\tau)$. We studied network dynamics using an effective action for wiggly cosmic strings, and introduced the averaged macroscopic equations into CMBact, allowing us to compute CMB anisotropies from these strings. CMBact has already built in the option to study wiggly strings, but this was done through a single constant parameter. Here, for the first time, we were able to take into account the time evolution of wiggles and their influence on the macroscopic equations of motion for the string network. This full treatment brought important changes in modelling wiggly cosmic string networks. From Fig.~\ref{fig:CMBwiggly} we see that wiggly strings can produce a lower signal in CMB anisotropy than ordinary strings (when the other parameters are fixed), which had not been appreciated before our work. We have also compared our analytic estimation~(\ref{Clwiggly}) to our numerical results from the CMBact code. The comparison shows that the main trend for $C_l$ (decreasing of $C_l$ as $\mu$ increases, for multiple moments $1<<l$) is captured correctly. We argue that for reliable constraints on wiggly string networks through the CMB signal, the evolution of string currents and its effect on string dynamics -- as captured by our wiggly model -- should be taken into account.  

We point out that comparing results from our analytic wiggly string network evolution and the standard VOS model for ordinary cosmic strings, has a broad resemblance to the differences that appear between Abelian-Higgs and Nambu-Goto numerical simulations for strings. In particular, increasing the amount of wiggles $\mu$ leads to slower rms velocities and a lower contribution to the string-induced CMB anisotropy decreases. This is similar to the difference between Abelian-Higgs and Nambu-Goto string networks, where the Abelian-Higgs strings tend to be slower and produce a lower CMB signal. This is, at present, a speculative observation requiring further investigation to see if a more firm analogy may be established.

The other type of strings that were scrutinized in this work are superconducting cosmic strings. It was shown that if we use the microscopic charge conservation~(\ref{ChargeCond1}) and the chiral condition ($\kappa, \Delta \rightarrow 0$, which appears in field theory studies of cosmic strings), we can obtain the averaged equations of motion~(\ref{ChEq1})-(\ref{ChEq3}) without specifying the precise dependence on string currents $f(\kappa_i,\Delta_i/\gamma)$ in the action~(\ref{ActionGeneral}). This implies that the debate on the correct form of the Lagrangian for superconducting strings~\cite{CarterPeter2} -- while important from a fundamental physics point of view -- does not have a crucial impact on phenomenological descriptions based on averaging the microscopic dynamics. By comparison to the work of~\cite{OliveiraAvgoustidisMartins}, we notice that the introduction of additional currents for superconducting strings only led to the change $s \rightarrow \beta s$ in the macroscopic VOS model. Introducing the appropriate modifications to CMBact, we have found that the string-induced CMB anisotropies tend to decrease with increasing the charge $Q$ of superconducting stings. Since the charge $Q$ does not have a scaling behaviour in the full range of physically relevant expansion rates~(\ref{Qscaling}), but generally decreases with evolution, the main effect on the CMB anisotropy comes from the initial charge $Q_0$ at the moment of string formation. We varied the initial charge to obtain a range of network dynamics histories and computed the corresponding CMB signal predictions. Numerical simulations are needed to further quantify the relevant model parameters.

The approach developed here can be useful in Markov chain Monte Carlo analysis of cosmological models with cosmic strings~\cite{Tom}. It allows to obtain more accurate constraints on wiggly and superconducting string network parameters directly from CMB observations. 

Finally, it is worth noting that the effects of the presence of currents on strings, described by our macroscopic VOS model, will also have a non-trivial impact on other observational windows for cosmic string networks, such as the stochastic gravitational wave background generated by string networks~\cite{Damour:2000wa,Damour:2001bk,Damour:2004kw,Binetruy:2009vt,Binetruy:2010cc,OCallaghan:2010rlo,OCallaghan:2010jrm,FigueroaHindmarshUrrestilla,AvelinoSousa2,KuroyanagiTakahashiYonemaruKumamoto}. Our results on string evolution and the methodology developed here for computing the two-point (unequal time) correlator will be useful for further studies in this direction too.

\begin{acknowledgments}

This work of IR is supported by the FCT fellowship (SFRH/BD/52699/2014), within the FCT PD Program PhD::SPACE (PD/00040/2012). IR is grateful for the hospitality of the University of Nottingham, where part of this work was carried out. The work of AA was partly supported by an Advanced Nottingham Research Fellowship at the University of Nottingham. CJM is supported by an FCT Research Professorship, contract reference IF/00064/2012, funded by FCT/MCTES (Portugal) and POPH/FSE (EC).  We would like to thank Patrick Peter, Tom Charnock, Jos\'e Pedro Vieira and colleagues from \href{http://polozov.cf/}{P.S.} for fruitful discussions and help. 

\end{acknowledgments}

\section*{Appendix: Analytic expressions for equal-time correlators} \label{Appendix}

As shown in~\cite{ACMS} the integral~(\ref{Correlator}) can be expanded in the following way

\begin{equation}
\begin{gathered} 
\left< \Theta^I(k,\tau_1) \Theta^J(k,\tau_2) \right> = \frac{f(\tau_1,\tau_2,\xi_0) \mu_0^2}{k^2(1-\upsilon^2)} \times \\
\sum_{i=1}^{6} A_i^{IJ} \left[ I_i(x_{-},\rho) - I_i(x_{+},\rho) \right], 
\end{gathered} 
\end{equation}
where $I$, $J$ correspond to the ``00", scalar, vector and tensor components of the stress-energy tensor and the form of the six integrals $I_i$ are as given in~\cite{ACMS}.

The coefficients $A_i^{IJ}$, together with the full expressions for the analytic equal time correlators $B^{IJ}$, are listed below (where, in this Appendix, we use the definitions $\rho = k | \tau_1 - \tau_2| \upsilon$, $x_{1,2}=k \xi_0 \tau_{1,2}$, $x_{\pm} = (x_1 \pm x_2)/2$):

\begin{align*}
 & A_1^{00-00} &=& &2 \hat{U}^2, \\
 & A_i^{00-00} &=& &0, \\
 &  &   &  &\qquad (i=2,..,6)  \\
 & A_1^{00-S} &=& &\hat{U}( \hat{T} + (2 \hat{U}- \hat{T} )v^2 ),  \\
 & A_2^{00-S} &=& &- 3 \hat{U} \left( \hat{T} (1-v^2) + \hat{U} v^2 \right)\\
 &A_3^{00-S} &=& &0\\
 &A_4^{00-S} &=& &-3 \hat{U}^2 v^2\\
 &A_5^{00-S} &=& &3 \hat{U}^2 v^2 \\
 &A_6^{00-S} &=& &0\\
   \end{align*}
\begin{align*}
 & A_1^{S-S} &=& &\frac{-27 \hat{U}^2 v^4 + \rho^2 (\hat{T} + (2 \hat{U} - \hat{T})v^2)^2}{2 \rho^2}\\
 & A_2^{S-S} &=& &\frac{3 \left( 9 \hat{U}^2 v^4 + \rho^2 \left( \hat{T}^2 (1-v^2)^2 - \hat{U}^2 v^4 \right)  \right) }{2 \rho^2} \\
 &A_3^{S-S} &=& &-\frac{9}{2} \left( \left( \hat{U} v^2 + \hat{T} (1-v^2)\right)^2 - 4 v^2 Q^2  \right) \\
 &A_4^{S-S} &=& &\frac{3 \hat{U} v^2 \left( 9 \hat{U} v^2 - \rho^2 (\hat{T}(1-v^2) + 2 \hat{U} v^2) \right)}{\rho^2} \\
 &A_5^{S-S} &=& &-\frac{3 \hat{U} v^2 \left(9 \hat{U} v^2 - \rho^2 (\hat{T}(1-v^2) + 2 \hat{U} v^2) \right)}{\rho^2} \\
 &A_6^{S-S} &=& &9 v^2 \left( \hat{U}^2 v^2 + \hat{T} \hat{U} (1-v^2) - 2 Q^2 ) \right)\\
 & A_1^{V-V} &=& &\frac{3 \hat{U}^2 v^4 + \rho^2 v^2 Q^2}{\rho^2} \\
 & A_2^{V-V} &=& &- \frac{3 \hat{U}^2 v^4}{\rho^2} \\
 &A_3^{V-V} &=& &\left( \hat{U} v^2 + \hat{T} (1-v^2) \right)^2 - 4 v^2 Q^2 \\
 &A_4^{V-V} &=& &-\left(6/\rho^2 - 1 \right) \hat{U}^2 v^4 - v^2 Q^2 \\
 &A_5^{V-V} &=& &\left(6/\rho^2 - 1 \right) \hat{U}^2 v^4 + v^2 Q^2 \\
 &A_6^{V-V} &=& &-2 v^2 \left(\hat{U}^2 v^2 + \hat{T} \hat{U} (1-v^2) - 2 Q^2 \right) \\
 & A_1^{T-T} &=& &\frac{\rho^2 \hat{T}^2 \left( 1-v^2 \right)^2 -3 \hat{U}^2 v^4}{4 \rho^2} \\
 & A_2^{T-T} &=& &\frac{3 \hat{U}^2 v^4 - \rho^2 \left( \hat{T}^2 \left( 1-v^2 \right)^2 - \hat{U}^2 v^4 \right) }{4 \rho^2} \\
 &A_3^{T-T} &=& &-\frac{1}{4} \left( \hat{U} v^2 + \hat{T}(1-v^2) \right)^2 + v^2 Q^2 \\
 &A_4^{T-T} &=& &\frac{v^2 \left( 3 \hat{U}^2 v^2 +\rho^2 \left( \hat{T} \hat{U}(1-v^2) + 2 Q^2 \right) \right)}{2 \rho^2} \\
 &A_5^{T-T} &=& &-\frac{v^2 \left( 3 \hat{U}^2 v^2 + \rho^2 \left( \hat{T} \hat{U}(1-v^2) + 2 Q^2 \right) \right)}{2 \rho^2} \\
 &A_6^{T-T} &=& &\frac{v^2}{2} \left( \hat{U}^2 v^2 + \hat{T} \hat{U} (1-v^2) - 2 Q^2 \right)
\end{align*} 
\begin{widetext}
\begin{eqnarray} 
   \label{B001Equal}
    & B^{00-00}(\tau) = 2 \hat{U}^2 (\cos(x)-1+x Si(x)), \nonumber\\
    \label{BS01Equal}
    & B^{00-S} = \frac{1}{2 x} \left( \hat{U}(2\hat{T}+v^2(\hat{U}-2\hat{T}))(x \cos(x) + 3 \sin(x)+x (x Si(x) - 4)) \right), \nonumber\\
    \label{BS1Equal}
    & B^{S-S} = \frac{1}{16 x^3} \biggl( \biggl[ 8 \hat{T} \hat{U} v^2 (1-v^2)(x^2-18)+8 \hat{T}^2 (1-v^2)^2 (x^2-18)+\hat{U}^2 v^4(11x^2-54)+288v^2 Q^2 \biggl]  x \cos(x) + \nonumber \\
    &+x^3 \biggl[ 32 \left( 3 v^2 Q^2 - \hat{U}^2 v^4 -\hat{T} \hat{U} v^2 (1-v^2)-\hat{T}^2 (1-v^2)^2) \right)+\left( 11 \hat{U}^2 v^4 + 8\hat{T} \hat{U} v^2 (1-v^2)+8 \hat{T}^2 (1-v^2)^2 \right) x Si(x) \biggl]- \nonumber
            \end{eqnarray}
    \begin{eqnarray} 
    &   - 3 \sin(x) \biggl[ 8 \hat{T} \hat{U} v^2 (1-v^2)(x^2-6)+8 \hat{T}^2 (1-v^2)^2(x^2-6)-\hat{U}^2 v^4 (18+z^2)+96 v^2 Q^2 \biggl], \nonumber \\
    \label{BV1Equal}    
    &  B^{V-V} = \frac{1}{24 x^3} \biggl( 3 x \cos(x) \left[ 16 \hat{T} (1-v^2) (\hat{T} - (\hat{T}-\hat{U})v^2) + \hat{U}^2 v^4 (6+z^2) + 4 v^2 (x^2-8) Q^2 \right]  + \nonumber \\
    & x^3 \left[ 16 \hat{T} (1-v^2)(\hat{T} - (\hat{T}-\hat{U})v^2) - 32 v^2 Q^2 + 3 v^2 x (\hat{U}^2 v^2 + 4 Q^2 ) Si(x) \right] - \nonumber \\
    & 3 \sin(x) \left[ 16 \hat{T} (1-v^2) (\hat{T}-(\hat{T}-\hat{U})v^2) + \hat{U}^2 v^4 (6-x^2) + 4 v^2 (x^2 - 8) Q^2  \right] \biggl), \nonumber\\
        \label{BT1Equal}    
    &  B^{T-T} = \frac{1}{96 x^3} \biggl( 3 x \cos(x) \left[ (3 \hat{U}^2 v^4 + 8 \hat{T} \hat{U} v^2 (1-v^2) +8 \hat{T}^2 (1-v^2)^2 ) (x^2 - 2) + 16 v^2 (2+x^2) Q^2 \right]  + \nonumber \\
    & + x^3 \left[ 64 \hat{T} (1-v^2) (v^2 (\hat{T}-\hat{U})-\hat{T})-64 v^2 Q^2 + 3 x (3 \hat{U}^2 v^4 + 8 \hat{T} \hat{U} v^2 (1-v^2) +8 \hat{T}^2 (1-v^2)^2 +16 v^2 Q^2 ) Si(x) \right] + \nonumber \\
    & + 3 \sin(x) \left[ \hat{U}^2 v^4 (6-5 x^2) + 8 \hat{T} \hat{U} v^2 (1-v^2) (2 + x^2) + 8 \hat{T}^2 (1-v^2)^2 (2+x^2)+16 v^2 (x^2 - 2) Q^2 \right] \biggl) \nonumber.
\end{eqnarray}
\end{widetext}

\bibliography{wiggly}
\end{document}